\documentclass[useAMS,usenatbib]{mn2e}
\usepackage{amsfonts}
\usepackage{bbm}
\usepackage{graphicx}
\usepackage{times}
\usepackage{graphicx}
\usepackage{rotating}
\usepackage{array}
\usepackage{xspace}
\usepackage[table]{xcolor}
\usepackage{amsmath}
\usepackage{amssymb}

\usepackage{amsmath}
\usepackage{natbib}

\title[orbit properties of prolate galaxies]
{
{Orbit properties of massive prolate galaxies in the Illustris simulation}
\author[Y. Wang et al.]
{Yougang Wang$^1$\thanks{E-mail:wangyg@bao.ac.cn}, Shude Mao$^{2,1,3}$, Hongyu Li$^{1,4}$, Dandan Xu$^{5}$, 
Xuelei Chen$^1$, \and Volker Springel$^{5,6,7}$ \\
$^1$Key Laboratory of Computational Astrophysics, National Astronomical Observatories,
Chinese Academy of Sciences, Beijing, 100012 China \\
$^2$Physics Department and Tsinghua Centre for Astrophysics, Tsinghua University, Beijing 100084, China\\
$^3$Jodrell Bank Centre for Astrophysics, School of Physics and Astronomy, The University of Manchester, Oxford Road,\\ Manchester M13 9PL, UK\\
$^4$University of Chinese Academy of Sciences, Beijing 100049, China\\
$^5$Heidelberg Institute for Theoretical Studies, Schloss-Wolfsbrunnenweg 35, 69118 Heidelberg, Germany\\
$^6$Zentrum f\"{u}r Astronomie der Universit\"{a}t Heidelberg, ARI, M\"onchhofstr. 12-14, 69120 Heidelberg, Germany\\
$^7$Max-Planck-Institute for Astrophysics, Karl-Schwarzschild-Str. 1, 85740 Garching, Germany}
}
\begin{document}

\date{Accepted . Received .}

\pagerange{\pageref{firstpage}--\pageref{lastpage}} \pubyear{2014}

\maketitle

\label{firstpage}

\begin{abstract}
We explore orbit properties of 35 prolate-triaxial galaxies selected from the Illustris cosmological hydrodynamic simulation. 
We present a detailed study of their orbit families, and also analyse relations between the relative abundance of the orbit families and the spin parameter, triaxiality, the ratio of the angular momentum and the baryon fraction.    
We find that box orbits dominate the orbit structure for most prolate-triaxial galaxies, especially in the central region. The fraction of irregular orbits in the prolate-triaxial galaxies is small, and it increases with galaxy radius.
Both the x-tube and z-tube orbits are important in prolate-triaxial galaxies, especially in the outer regions. The fraction of box orbits for prolate-triaxial galaxies ($0.7<T<1$) decreases with the triaxiality of the stars, while the fraction of x-tube orbits increases with the triaxiality for a given axis ratio.
The fraction of box orbits increases and the fractions of x-tube and z-tube orbits weakly decrease with increasing baryon fraction. These results help to understand the structure of prolate-triaxial galaxies and provide cross-checks for constructing dynamical models of prolate-triaxial galaxies by using the Schwarzschild or the Made-to-Measure methods. 
  
\end{abstract}


\section{Introduction}

The misalignment between the kinematic and photometric axes shows the complex structure of
elliptical galaxies \citep[e.g.][]{1991ApJ...383..112F}. The brightness profiles of ellipticals indicate that their intrinsic shapes could be oblate, prolate, or triaxial, which is also 
supported by theoretical studies \citep[e.g.][]{1985MNRAS.212..767B,1996ApJ...460..136M,2008ApJ...677.1033W,2017ApJ...844..130W} and by N-body simulations \citep[e.g.][]{2002ApJ...574..538J}.
Kinematics shows that most rotating elliptical galaxies are either oblate or slightly triaxial \citep{2011MNRAS.414.2923K,2014MNRAS.444.3340W,2015MNRAS.454.2050F,2016ARA&A..54..597C}. 
However, some elliptical galaxies show minor-axis rotation (prolate rotation, i.e., the rotation is about the photometric major axis), such as NGC 1052 \citep{1979ApJ...229..472S,1986ApJ...303L..45D}, NGC 4406, NGC 5982, NGC 7052, NGC 4365, NGC 5485 
\citep{1988A&A...195L...5W}, NGC 3923 \citep{1998MNRAS.294..182C}, M87 \citep{1988ApJS...68..409D,2014MNRAS.445L..79E}, NGC 4473 \citep{2013MNRAS.435.3587F}; eight elliptical galaxies from the
CALIFA survey \citep{2017A&A...606A..62T} also show such rotation. Moreover, stellar kinematics from two-dimensional integral field unit (IFU) observations indicate that prolate rotation often coexists with oblate rotation in many elliptical galaxies  \citep[e.g.][]{2006MNRAS.373..906M,2011MNRAS.414.2923K,2014MNRAS.445L..79E}.   

It is important to reveal the formation scenario of the `prolate' rotation of elliptical galaxies. The dynamical stability and orbit properties of prolate rotation has been extensively studied  \citep[e.g.][]{2011ApJ...728..128D,2014A&A...563A..19Z}. However, the formation of prolate rotation in elliptical galaxies is still unclear. From simulations, \cite{2015MNRAS.449...49R} found that prolate systems with minor axis rotation could be produced by the merger of two equal-mass disc galaxies.  Similar results have been obtained for gas-poor mergers in a hydrodynamic simulation by \cite{2014MNRAS.444.1475M}. Recently, \cite{2017A&A...606A..62T} found that prolate galaxies could be the results of dry polar mergers, and the amplitude of prolate rotation depends on the initial bulge-to-total stellar mass ratio of its progenitor galaxies. \cite{2017ApJ...850..144E} identified 59 prolate rotators in the Illustris cosmological hydrodynamic simulation and found that the emergence of prolate rotation is strongly correlated with the time of their last significant merger.
Also, using the Illustris simulation, \citet[hereafter Li2018]{2018MNRAS.473.1489L} found 35 out of a total of 839 galaxies with stellar mass larger than $10^{11}M_{\odot}$ are prolate-triaxial, and these prolate-triaxial galaxies are formed by major dry mergers. All studies show that galaxies with prolate rotation are most likely formed by dry major mergers, and the number of these objects may dominate at the high-mass end.

Orbits are the fundamental building blocks of galaxies and therefore their properties greatly affect their internal structures. In this paper, we study the orbit properties of the prolate-triaxial galaxies selected in Li2018, which can help us to understand the dynamics of these systems.       
Moreover, the orbit families have important implications for constructing dynamical models with the Schwarzschild orbit superposition method \citep{1979ApJ...232..236S,2006MNRAS.366.1126C,2012MNRAS.427.1429W,2013MNRAS.435.3437W}, Made-to-Measure method \citep{1996MNRAS.282..223S,2012MNRAS.421.2580L,2013MNRAS.428.3478L,2014ApJ...792...59Z} and the Torus method \citep{2008MNRAS.390..429M,2017MNRAS.470.2949W}. Compared with previous theoretical studies, our prolate-triaxial galaxies are taken from the state-of-the-art Illustris simulation, hence the model itself is self-consistent.

The structure of the paper is as follows. In Section~\ref{sec:simulation},  we introduce the simulation data and the selected prolate-triaxial galaxies.  In Section~\ref{sec:orbit_integration}, we detail the orbit integration, and in Section~\ref{sec:orbit_cla} we explain our orbit classification. In Section ~\ref{sec:orbit_pro}, we present our main results of the orbit properties. Our summary and conclusions are given in Section ~\ref{sec:summary}.   




\section{simulations and prolate galaxies}
\label{sec:simulation}
The Illustris project is a  large cosmological simulation of galaxy formation \citep{2014MNRAS.445..175G, 2014MNRAS.444.1518V}, which comprises
a suite of $N$-body/hydrodynamical simulations carried out with the moving mesh code \textsc{arepo} \citep{2010MNRAS.401..791S}.  
The simulation adopts a comprehensive set of physical models for galaxy formation, which can reproduce various observational constraints at different redshift, such as star formation rate density, galaxy gas fraction, mass-size relation \citep{2017MNRAS.469.1824X}, galaxy luminosity function, galaxy morphologies \citep{2014MNRAS.445..175G}, Tully-Fisher relation etc.     

The prolate-triaxial galaxies in this work are from Li2018, which are selected from the largest simulation (Illustris-1) of the Illustris project. Illustris-1 contains $1820^3$ dark matter particles and approximately $1820^3$ gas cells or stellar particles 
in a $\rm (106.5\ Mpc)^3$ box. The simulation evolves from $z=127$ to $z=0$ in a standard $\rm{\Lambda}$ cold dark matter cosmology with $\Omega_{\Lambda}=0.7274$,  $\Omega_m=0.2726$, $\Omega_b=0.0456$, $\sigma_8=0.809$, $n_s=0.963$ and $H_0=70.4\ \rm{km^{-1}Mpc^{-1}}$.  The details of the Illustris simulation can be found in \cite{2014MNRAS.445..175G}.

In Li2018, the galaxies have been selected from  Illustris-1 at redshift $z=0$ (snapshot 135) by stellar mass $M_{\star}>10^{11}M_{\odot}$ and light profile S{\'e}rsic index
 $n_{S\acute{e}rsic}>2$ \citep{1963BAAA....6...41S}.  The number of selected galaxies is 839. Each galaxy is assumed to be an ellipsoid with axis lengths $a\ge b\ge c$. The axis ratios $p=b/a$ and $q=c/a$ are measured from the stellar particles using the reduced inertia tensor method \citep{2006MNRAS.367.1781A}. The tensor is defined as
\begin{equation}
I_{ij}=\frac{\sum_km_kx_{k,i}x_{k,j}/r_k^2 }{\sum_km_k},\ \ i=1, 2, 3,\  j=1, 2, 3, 
\end{equation}
where $x_{k,1}$= $x_k$, $x_{k,2}=y_k$ and $x_{k,3}=z_k$ are the positions of the k-$th$ particle in the simulation. $m_k$ is the mass of the k-$th$ particle and $r_k=\sqrt{x_k^2+y_k^2/p^2+z_k^2/q^2}$ is the elliptical distance measured from the centre of the galaxy to the $k$-th particle.  

A prolate-triaxial galaxy satisfies $b/a-c/a<0.2$ 
and $b/a<0.8$. A total of 35 prolate-triaxial galaxies have been selected in Li2018 and the properties of all the prolate-triaxial galaxies are listed in Table B1 of Li2018. Some properties of the prolate-triaxial galaxies are also shown in Table~\ref{table:of_all}, such as $M_{c200} $, which is defined as the total mass enclosed in a sphere whose mean density is 200 times the critical density of the Universe.    
Among these 35 prolate-triaxial galaxies, we find that the axis ratio evolves no more than $15\%$ from $z=0.1258$ to $z=0$ for 28 galaxies. For the remaining seven galaxies (subhalo129771, subhalo177128, subhalo245939, subhalo277529, subhalo289892 and subhalo294574), the axis ratio evolves more than $15\%$ but less than $40\%$. This indicates that most galaxies in our sample are stable. The upper limit of the tumbling rate is $\rm {4.0\ km\ s^{-1}\ kpc^{-1}}$, which corresponds to a period of 1.5 Gyr, much longer than typical dynamical times in the central part of galaxies.

We also select six oblate-triaxial galaxies and six triaxial galaxies from the Illustris simulation for comparison purposes.  
These six oblate-triaxial galaxies and six triaxial galaxies are within the same mass range and have similar halo spin magnitudes as their prolate-triaxial counterparts, which are given in the middle and bottom sections in Table~\ref{table:of_all}.        

\begin{figure}
\includegraphics[angle=0, width=100mm]{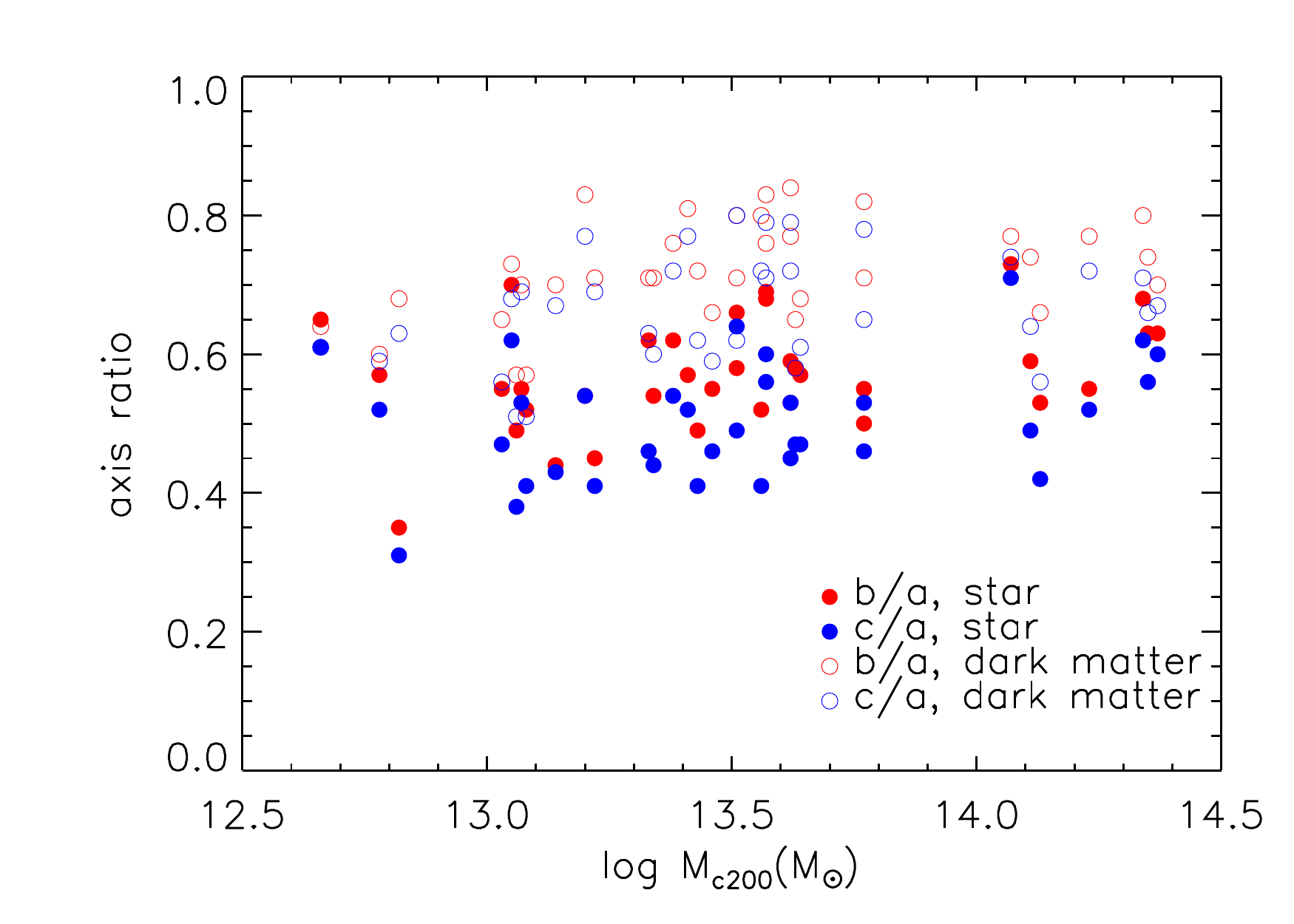}
\caption{Axis ratios $b/a$ (red) and $c/a$ (blue) calculated within the half stellar mass radius. The filled and open circles represent the results for the star and dark matter particles, respectively. The x-axis is the total mass enclosed within a sphere whose mean density is 200 times the critical density of the Universe.}
\label{fig:axis}
\end{figure}

Figure~\ref{fig:axis} shows the relation between $M_{c200}$ and the  axis ratios $b/a$ (red) and $c/a$ (blue)  calculated by the particles within the half stellar mass radius for both the stellar and dark mater particles for prolate-triaxial galaxies.  It is seen that  the shape of the dark matter halo is more spherical than that of the stars except for the galaxy 324170 (See Table~\ref{table:of_all}). 

\begin{figure}
\includegraphics[angle=0, width=100mm]{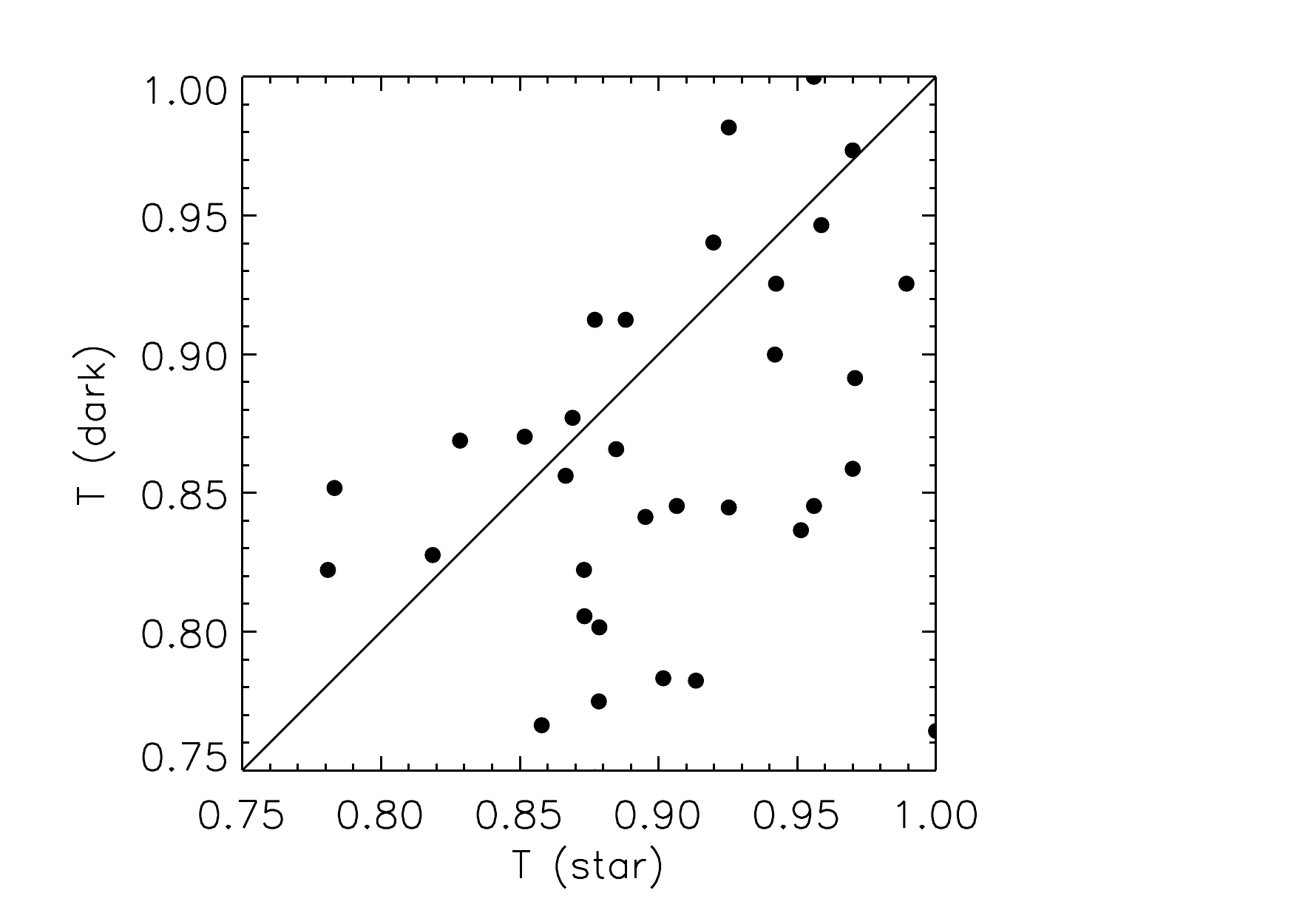}
\caption{Comparison of the triaxiality of the stars and that of the dark matter halo within the half stellar mass radius. These two are correlated, but have large scatter.}
\label{fig:triaxiality}
\end{figure}

In Figure~\ref{fig:triaxiality}, we show a comparison of the triaxiality of the stars with that of the dark matter halo within the half stellar mass radius for prolate-triaxial galaxies. The triaxiality  parameter $T$ is defined as $T=(a^2-b^2)/(a^2-c^2)$;  $T=1$ and $T=0$ for prolate and oblate systems, respectively. It is seen that  the triaxialities measured for the stars and the halo are both larger than 0.7, which again shows our selected galaxies are prolate/triaxial. And the shape of stars is different from that of the dark halo for most galaxies. The triaxialities of all galaxies are given in Table~\ref{table:comp}. It can be seen that for most prolate-triaxial galaxies, $T\gtrsim0.8$, while for oblate-triaxial galaxies, $T\leqslant0.3$. The ``triaxial" galaxies are somewhere in between.

The spin parameter in each galaxy is defined as \citep{2001ApJ...555..240B}
 \begin{equation}
 \lambda^{\prime}=\frac{L}{\sqrt{2}MVR},
 \end{equation}  
 where $L$ is the angular momentum within a sphere of radius  $R$ containing mass $M$. The galaxy circular velocity $V$ is defined as $V=\sqrt{GM/R}$. The format of the spin parameter we adopt here is different from the traditional one \citep{1979MNRAS.186..133E,2010gfe..book.....M}, which needs  to calculate the energy of the system to define the spin parameter.

 Columns (2)-(4) in Table~\ref{table:spin} show the spin parameters for stars ($\lambda^{\prime}_s$), dark matter ($\lambda^{\prime}_d$) and both the star and dark matter ($\lambda^{\prime}$) within the half stellar mass radius, respectively. It is seen that the spin parameters for both the stellar and dark matter components are small ($<0.07$). We also find, not surprisingly, that the spin parameters for stars in the six oblate-triaxial galaxies are larger than those in the prolate-triaxial galaxies and the triaxial galaxies.    
 
Column (5) in Table~\ref{table:spin} shows the angle ($\theta_L$) between the direction of the angular momentum of the dark matter and that of the stars.  We find that $\theta_L$ in 12 prolate-triaxial galaxies is larger than $45^{\circ}$, and $\theta_L$ in two galaxies (245939 and 294574) is close to $90^{\circ}$. The reason why these galaxies have such a large value of $\theta_L$ is not clear, we will return to this in a future work. For the six oblate-triaxial galaxies, all values of $\theta_L$ are smaller than $20^{\circ}$. For the six triaxial galaxies, one $\theta_L$ is larger than $45^{\circ}$.  

\begin{figure}
\includegraphics[angle=0, width=80mm]{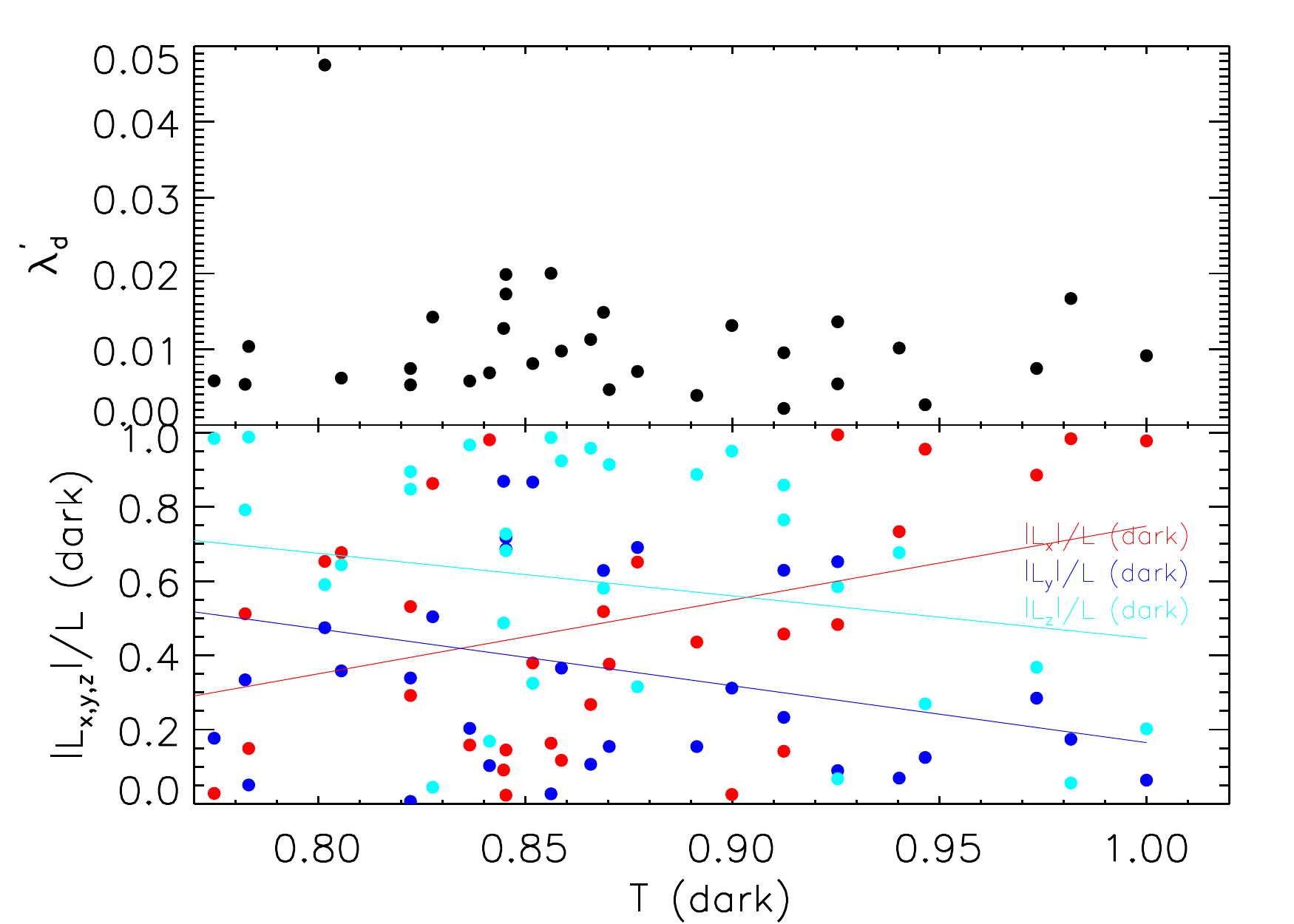}
\caption{Top: Relation between the triaxiality $T$ of the dark matter and the spin parameter $\lambda_d^{\prime}$  measured within the half stellar mass radius for prolate-triaxial galaxies. Bottom:  Relation between the  triaxiality of the dark matter halo and the fraction of $\left|L_x\right|/L$ (red), $\left|L_y\right|/L$ (blue) and $\left|L_z\right|/L$ (cyan) within the half stellar mass radius for prolate-triaxial galaxies. The red, blue and cyan lines are fitted straight lines with $Y=A+BX$ for the cases $\left|L_x\right|/L$,  $\left|L_y\right|/L$ and  $\left|L_z\right|/L$, respectively. }
\label{fig:L2}
\end{figure}

The upper panel of Figure~\ref{fig:L2} shows the relation between the triaxiality of the dark matter and the spin parameter measured within the half stellar mass radius for the prolate-triaxial galaxies. We find that there is no clear relation between the halo triaxiality and the spin parameter of the dark matter halo. The bottom panel of this figure shows the relation between the triaxiality  of the dark matter halo and the fraction of $\left|L_x\right|/L$ (red), $\left|L_y\right|/L$ (blue) and $\left|L_z\right|/L$ (green) within the half stellar mass radius.
Here, $L_x$, $L_y$ and $L_z$ are the angular momenta along the major, middle and minor axes of the system, respectively, and $L=\sqrt{L_x^2+L_y^2+L_z^2}$ is the total angular momentum.  It is noted that $L_x$ dominates the total angular momentum if the galaxies are highly prolate ($T>0.95$). Moreover, we find that there is a weak correlation between the  triaxiality of the dark matter halo and the fraction of  $\left|L_x\right|/L$ (the Pearson correlation is 0.43), while there is a weak anti-correlation between the triaxiality of the dark matter halo and the fraction of  
$\left|L_y\right|/L$ and  $\left|L_z\right|/L$. The Pearson correlation coefficients are -0.37 and -0.26 for  $\left|L_y\right|/L$ and  $\left|L_z\right|/L$, respectively.

\section{orbit integration}
\label{sec:orbit_integration}
The mass resolution of the Illustris simulation is quite high. Therefore, the particle number in each galaxy is large, and we randomly select one tenth of the star particles in each prolate-triaxial galaxy as the initial conditions for the orbit integration.  
The orbit number of each galaxy is presented in column (3) of Table~\ref{table:of_all}.    

In order to obtain the potential and the accelerations of the disk
particles, we follow  the self-consistent field (SCF) method
\citep{1992ApJ...386..375H}, more specifically, we use the code SCF.f directly.


The key point of the SCF method is to obtain an estimate of the mean
gravitational field by expanding the density and potential into a
set of simple orthogonal basis of potential-density pairs in
spherical coordinates. The density and potential are given as
\begin{equation}
\rho(r,\theta,\phi)=\sum_{n,l,m}A_{nlm}\rho_{nl}Y_{lm}(\theta,\phi),
\end{equation}
\begin{equation}\label{eq:pot}
\Phi(r,\theta,\phi)=\sum_{n,l,m}B_{nlm}\Phi_{nl}Y_{lm}(\theta,\phi),
\end{equation}
where $n$ is the radial expansion index and $l$ and $m$ designate the
angular terms. The force can be derived directly from the potential. In order to obtain a high accuracy for the force calculation, we adopt as maximum values for the radial expansion terms $n_{\rm max}=16$ and for the angular terms $l_{\rm max}=12$, which can give high accuracy for the force and yet achieve a fast speed for the orbit integration. 
The orbit integration is preformed with the DOP853 algorithm \citep{hai93}.  Each orbit is integrated for 500 dynamical times.  The dynamical time $T_{D}$ is defined as $T_D=2\pi R/V_c$, where $R$ is the radius of the particle $R=(x^2+y^2)^{1/2}$ and 
$V_c$ is the circular velocity of the particle. The circular velocity is defined as $V_c=\sqrt{|xa_x+ya_y|}$, $a_x$ and $a_y$ are the forces along the $x$ and $y$-axes, respectively.   
  
\section{orbit classification} 
\label{sec:orbit_cla}
To classify the orbits, we use the spectral classification routine of \citet[hereafter CA98]{1998MNRAS.298....1C}. We will refer to this method as CA.
The key point of CA98 is to find the number of fundamental frequencies.  For a three dimensional system, the regular orbits have no more than three fundamental frequencies, while the irregular orbits have more than three fundamental frequencies. 
We use the latest version of \textsc{taxon.for} to classify the orbits.
Compared with its original form, the new version of the code uses the Frequency Modified Fourier Transform ~\citep[FMFT]{1996CeMDA..65..137S} to extract lines,  and the spectral
analysis is performed on both the position and velocity components  $X(t)+iV(t)$  (\citealt{2016MNRAS.463.3499W}, hereafter Wang2016).  

\begin{figure*}
\begin{center}
\includegraphics[height=0.355\textwidth]{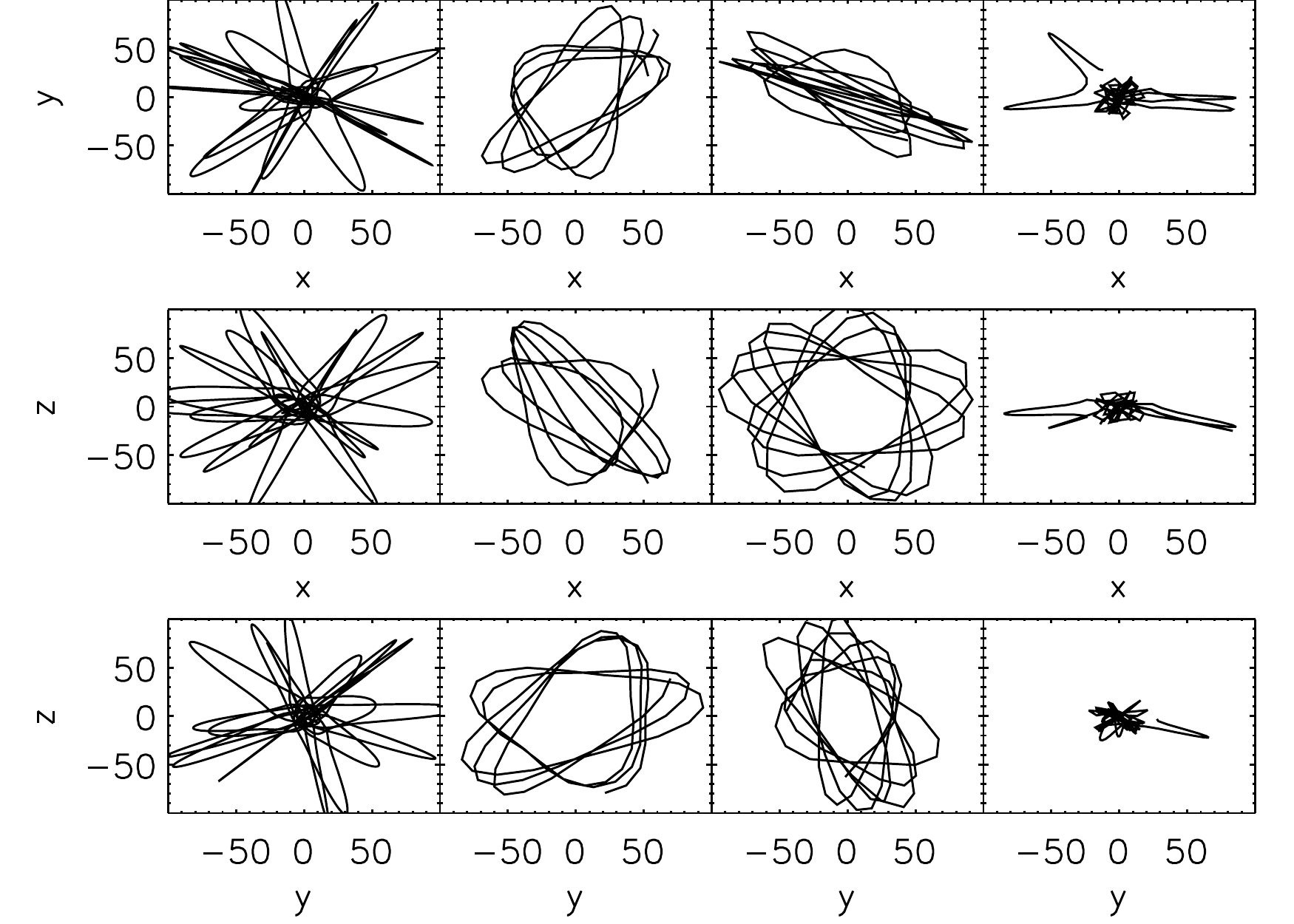}
\includegraphics[height=0.355\textwidth]{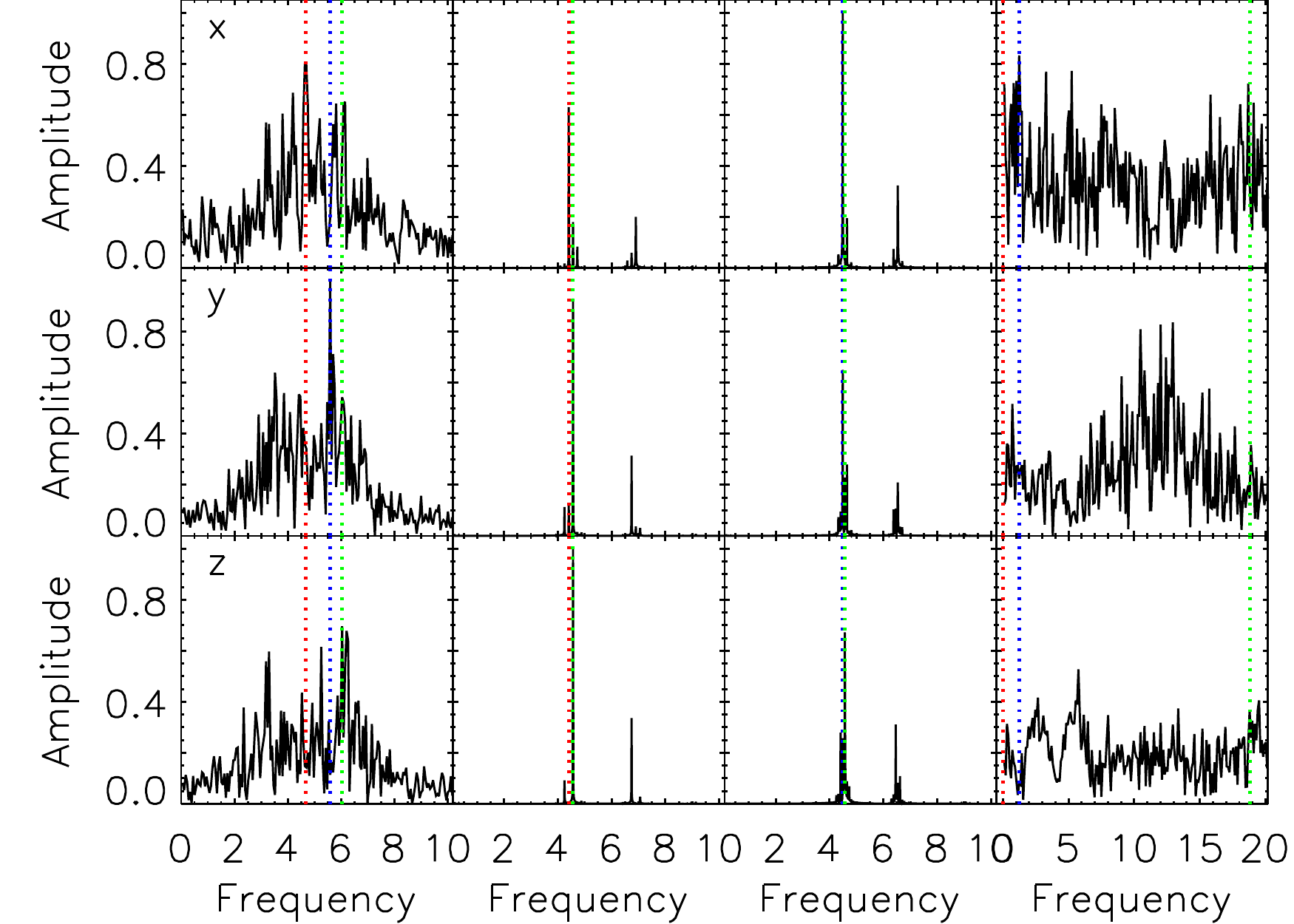}
\caption{Left panels: Orbit samples in subhalo 135289. From left to right: box, x-tube, z-tube and irregular orbits. Right panels: Frequency spectra for the corresponding orbits. 
From top to bottom, we show the results for the z, y and x components, respectively. The red, blue and green sold lines denote the position of the main frequencies $\omega_x$, $\omega_y$ and $\omega_z$. 
From left to right, $[\omega_x,\omega_y,\omega_z]$=[4.68, 5.59, 6.04], [4.40, 4.56, 4.56],[4.49, 4.49, 4.57], [0.01,1.25,18.82].   }
\end{center}
\label{fig:orbit}
\end{figure*}

In CA98, orbits are classified into box, x-tube, y-tube, z-tube and irregular families. The former four are regular orbits.   Box orbits show no sense of rotation and the orbits  can cross close to the potential center. Tube orbits tend to rotate around the centre of the system, the x-, y- and z-tube orbits refer to the orientation of the orbits being along the major, intermediate and  minor axes of the system. If the main frequency in each component is $\omega_i$, then a resonance is defined as
\begin{equation}
l_1\omega_x+l_2\omega_y+l_3\omega_z=0
\end{equation}      
for non-trivial combinations of integers $l_1$, $l_2$ and $l_3$. For the box orbits, the main frequencies are incommensurable. An orbit is taken as an x-tube if  the main frequencies of $\omega_z$ and $\omega_y$ show a 1:1 resonance, i.e., $l_2:l_3=1:1$ with $l_1$ being arbitrary. An orbit is classified as a z-tube  if  $\omega_x$ and $\omega_y$ show a 1:1 resonance ($l_1$:$l_2$=1:1).  The y-tube orbits are unstable and rare in a three dimensional system \citep[e.g.][]{2008gady.book.....B,1996ApJ...460..136M}, which shows a 1:1 resonance for $\omega_x$ and $\omega_z$. Figure~\ref{fig:orbit} shows four orbit examples (left) and their corresponding Fourier spectra (right).

The orbit fractions for box, x-tube, y-tube, z-tube and irregular families are given in columns (7)-(11) of Table~\ref{table:of_all} for all stars (the star within the half stellar mass radius). It is seen that the box families dominate for the most prolate-triaxial galaxies, and the fraction of box orbits increases if only the orbits within the half stellar mass are considered. The y-tube and irregular orbit fractions are small. It is known that there are no stable tube orbits around the intermediate axis in a triaxial system for a non-separable potential with a central density core (obtained numerically by \citealt{1979ApJ...233..872H}) and for all separable systems ~\citep{1984PhDT........87D,1985MNRAS.216..273D}, therefore, it is reasonable that we have obtained a small fraction for the y-tube orbits in the simulated galaxies. It is also noted that the z-tubes are important in prolate-triaxial systems, such as in  subhalo 138413, 16937 and 30430, which is different from previous studies \citep[e.g.][]{2016ApJ...818..141V}.  The sum of the box and z-tube orbits dominates in the oblate-triaxial and prolate-triaxial galaxies, whereas the fractions of  x-tube orbits in both oblate-triaxial and triaxial galaxies are small. Generally, the long-axis tube orbits include inner long-axis tube orbits and the outer long-axis tube orbits ~\citep{1985MNRAS.216..273D}. It is difficult to distinguish the inner and outer long-axis tube orbits using the spectral analysis, therefore, we only consider the population of long-axis tube orbits as a whole here.

For comparison, we also use another method to classify the orbits, which is based on the values of the three angular momentum components. We will hereafter refer to this method as AM. The detailed description of the AM method is given in Appendix~\ref{AppendixB}. In Table~\ref{table:comp}, we show the orbit population for both methods. It is seen that the AM method gives more box and irregular orbits than the CA routine. If we compare the orbit population one by one between the CA and AM methods, we find that they agree with each other for $80\%$ of the orbits. A detailed comparison between two different orbit classifications is complex~\citep{2016MNRAS.463.3499W}, and is beyond the focus of the present paper. We only consider the results from the CA method in the following.    
  
\section{orbit properties}
\label{sec:orbit_pro}
\begin{figure}
\includegraphics[angle=0,width=120mm]{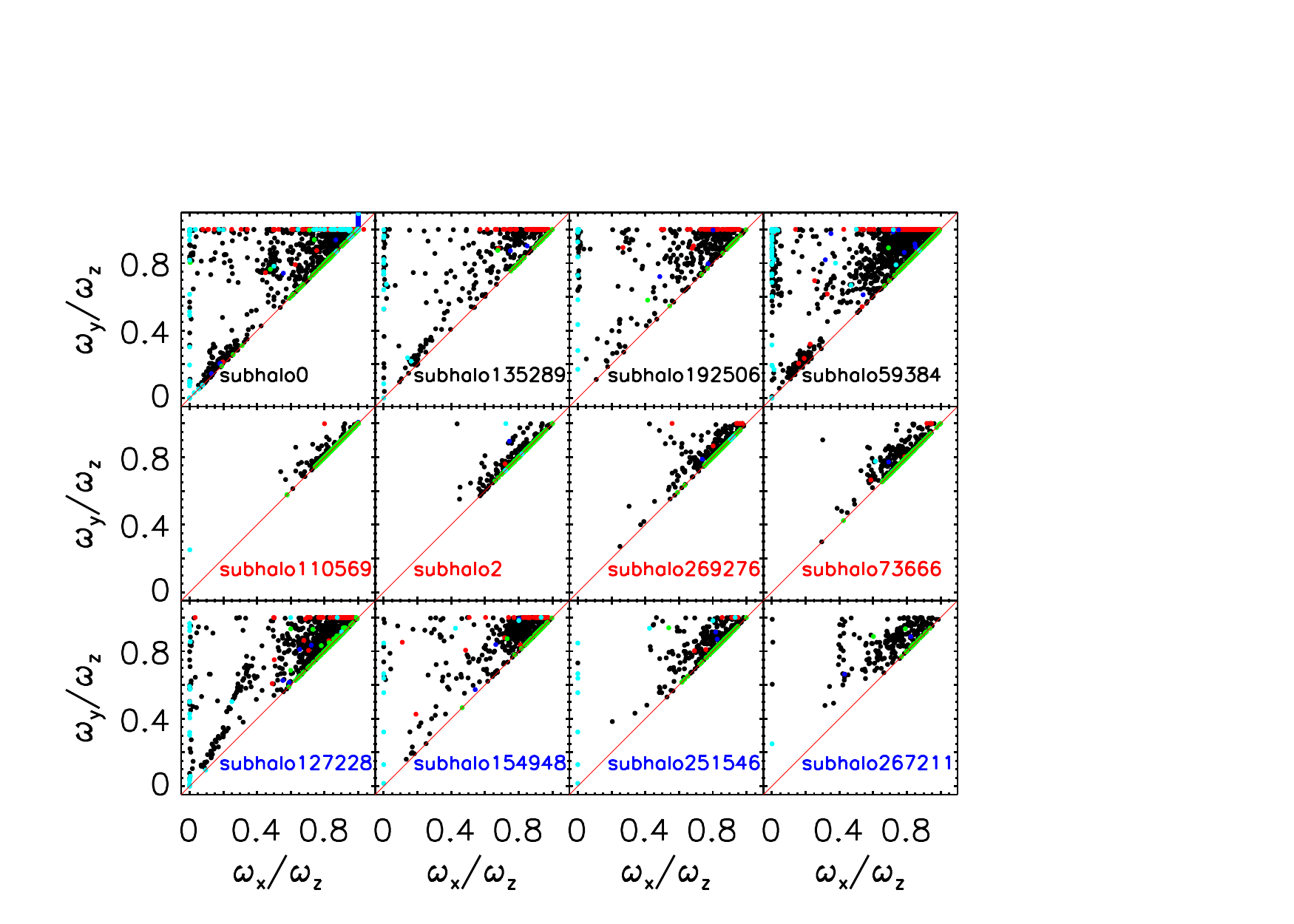}
\caption{Frequency maps in the plane of $\omega_x/\omega_z$ and $\omega_y/\omega_z$ for 4 prolate-triaxial galaxies (top panels with black labels),  4 oblate-triaxial galaxies (middle panels with red labels) and 4 triaxial galaxies (bottom panels with blue labels). In each panel, the black, red, blue, green and cyan points are the results for the box, x-tube, y-tube, z-tube and irregular orbits, respectively. The pink line in each panel is the diagonal line. Only $1\%$ randomly selected orbits are shown. } 
\label{fig:fre_map}
\end{figure}

In this section, we give a more detailed analysis of the orbit properties in the prolate-triaxial galaxies. In Figure~\ref{fig:fre_map}, we show frequency maps in the plane of $\omega_x/\omega_z$ and $\omega_y/\omega_z$ for  4 prolate-triaxial, 4 oblate-triaxial and 4 triaxial galaxies, which are selected randomly. Here $\omega_x$, $\omega_y$ and $\omega_z$ are three main frequencies from the spectrum of the $x$, $y$ and $z$ components, respectively. In CA98, the main frequency from each component should yield $\omega_x<\omega_y<\omega_z$ if x, y and z  are the major, intermediate and minor axes of the system. Therefore, the bottom right part in each panel is blank. For some galaxies, the main frequency ratios cross the diagonal line which is because the axes ratio of the halo  changes with the radius (See Li2018). It is also noted that the orbits with $\omega_x:\omega_y:\omega_z=1:1:1$ have a significant contribution in the orbit families. For both the prolate-triaxial and triaxial galaxies,  the main frequency ratios are distributed widely. For the oblate-triaxial galaxies, most main frequencies are distributed in a narrow range along the diagonal line. 

\begin{figure}
\includegraphics[angle=0, width=80mm]{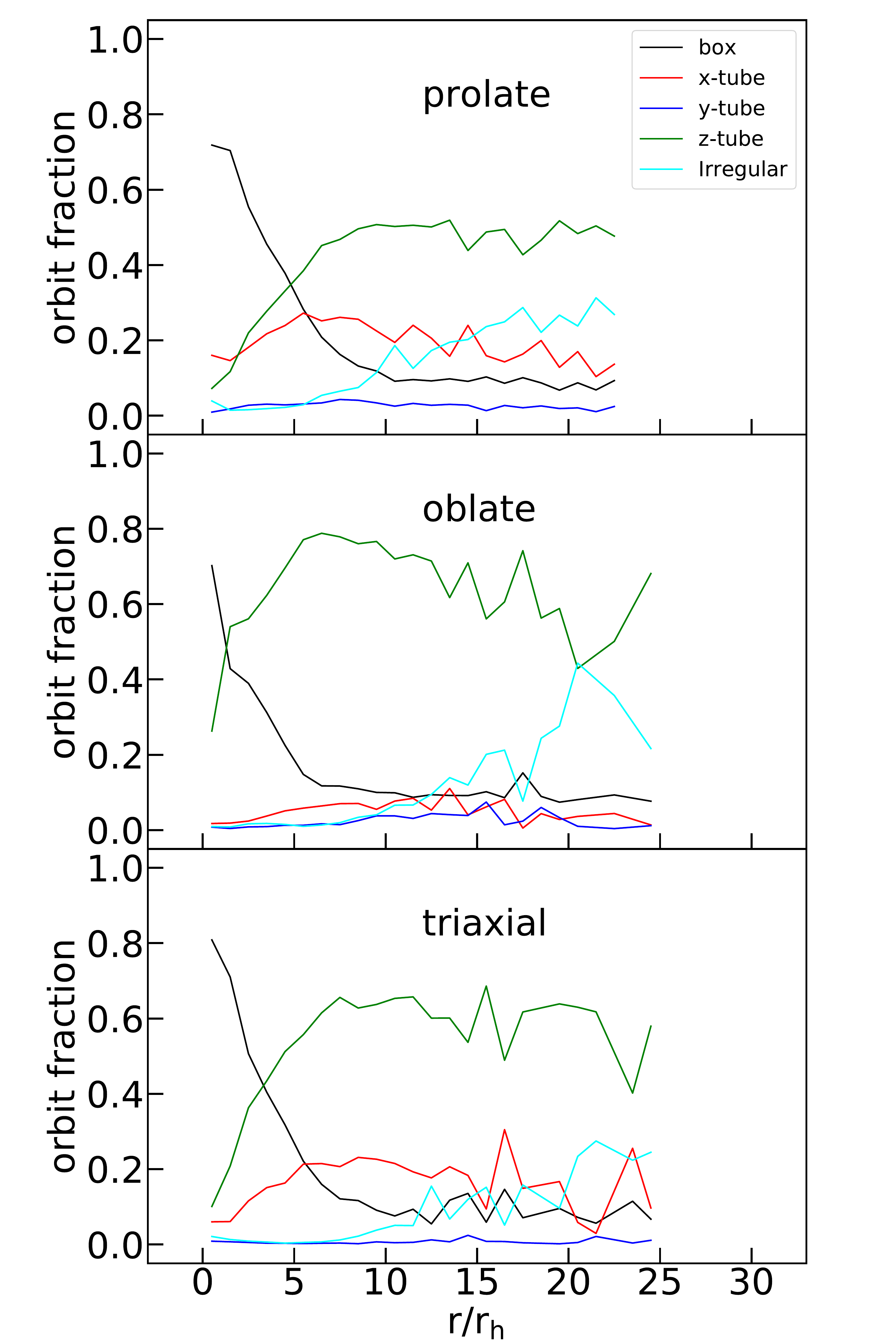}
\caption{Dependence of the average fraction for different orbit families on the galaxy radius (in units of the half-mass radius, $r_h$) for the prolate-triaxial (top), oblate-triaxial (middle) and triaxial (bottom) galaxies, respectively. The black, red, blue, green and cyan lines represent the results for the box, x-tube, y-tube, z-tube and irregular orbits, respectively.}
\label{fig:type_radius}
\end{figure}

Figure~\ref{fig:type_radius} shows the dependence of the average fraction for different orbit families as a function of the galaxy radius. It is seen that box orbits dominate in the central region for all galaxies, and the fraction of the box orbits decreases with the galaxy radius.  
The fraction of irregular orbits increases with the galaxy radius.
It is also found that the z-tube orbits dominate in the prolate-triaxial, oblate-triaxial and triaxial galaxies if the galaxy radius is larger than 5.5, 1.5 and  3.5 $r_h$, respectively. Although the z-tube orbits dominate in the outer region of the galaxy, they are different in the prolate-triaxial, oblate-triaxial and triaxial galaxies. The fraction of z-tube orbits in oblate-triaxial galaxies is larger than that in triaxial galaxies, and that in triaxial galaxies is larger than that in prolate-triaxial galaxies.

\begin{figure}
\includegraphics[angle=0, width=80mm]{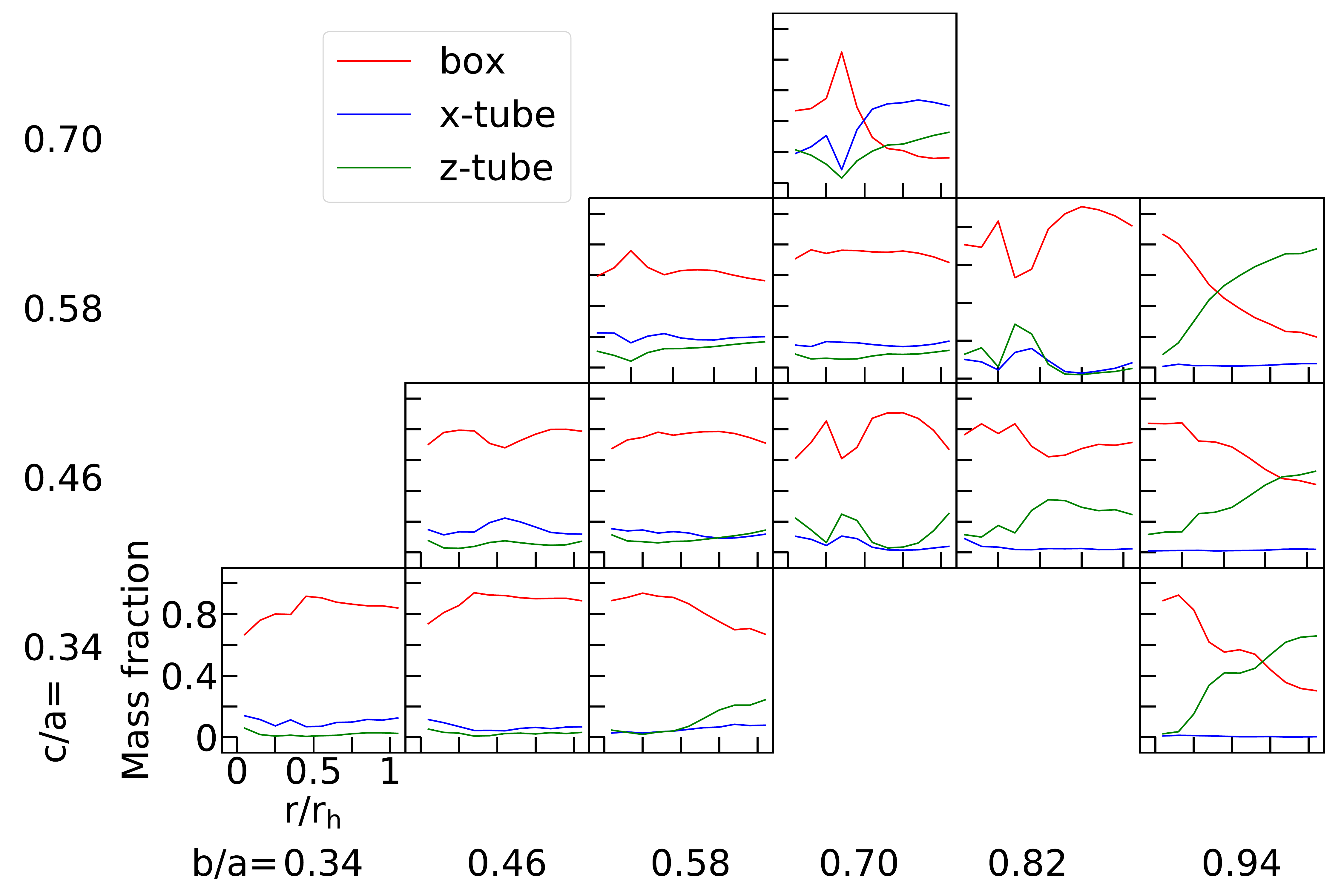}
\caption{Mass fractions of the box orbits (red line), x-tube orbits (blue lines) and the z-tube orbits (blue lines) as a function of axial ratio and radius for all galaxies in our sample. In each panel, the x-axis is the radius $r/r_h$ and y-axis is the mass fraction. The x-scale and y-scale run linearly from 0 to 1. Limited by the number of galaxies in our samples, some panels are missing. Galaxies in right panels are oblate galaxies, while those on the diagonal panels are prolate galaxies.}
\label{fig:mass_type}
\end{figure}

Figure~\ref{fig:mass_type} displays  the dependence of the fractions of mass from box, x-tube and z-tube orbits on the axial ratio and radius for all galaxies in our sample. It is seen that box orbits dominate the mass for prolate-triaxial systems. The mass contributed by the z-tube orbits increases with the galaxy radius when a galaxy is close to being oblate-triaxial. In prolate-triaxial systems, the mass from the x-tube orbits is larger than that from the z-tube orbits within $r_h$.

\begin{figure}
\includegraphics[angle=0, width=80mm]{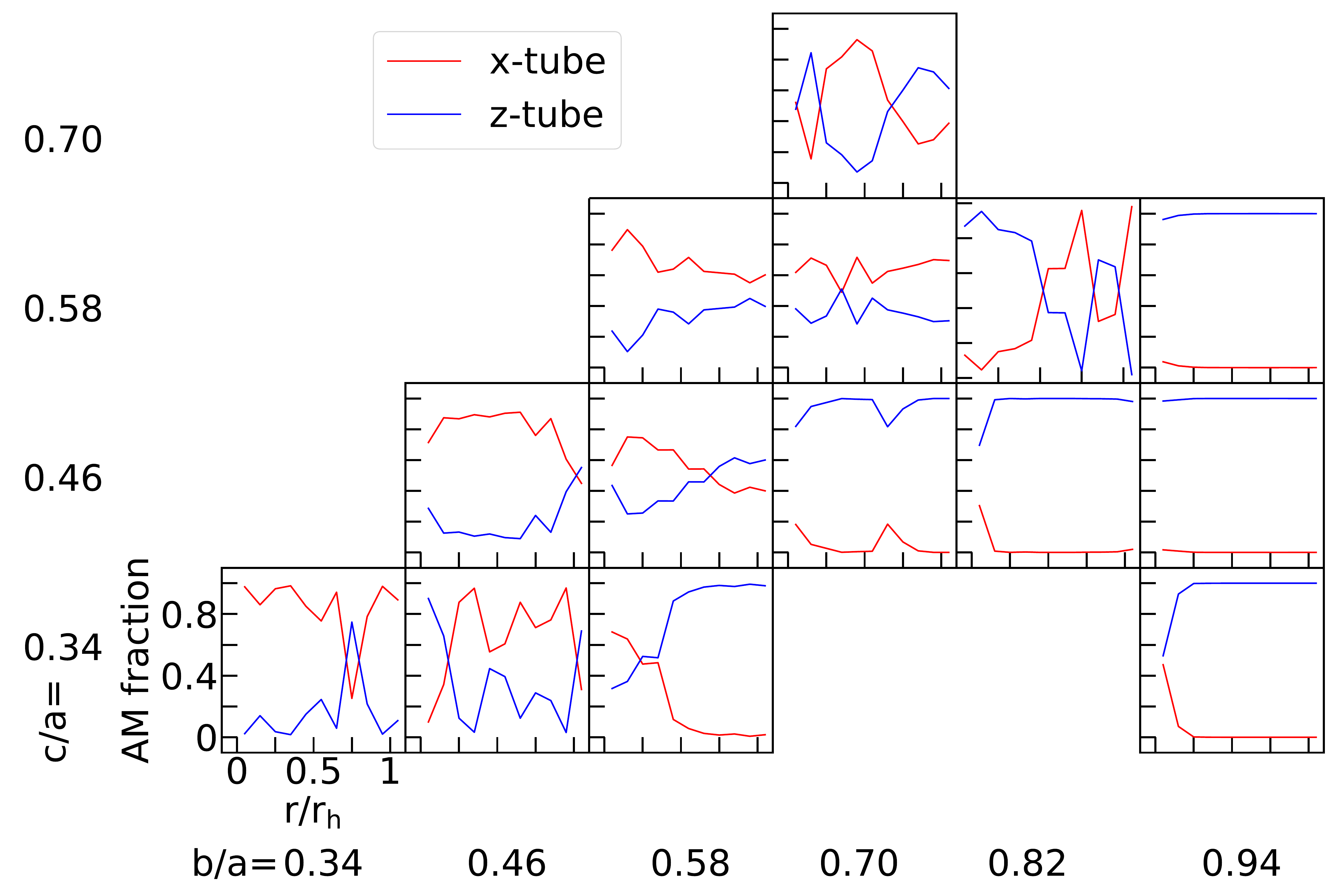}
\caption{Angular momentum (AM) fractions of the x-tube orbits (red lines) and the z-tube orbits (blue lines) as a function of axial ratio and radius for all galaxies in our sample. The angular momentum fractions of the x-tube and z-tube orbits are defined as $L_{\rm x-tube}^2/L_{\rm total}^2$ and $L_{\rm z-tube}^2/L_{\rm total}^2$, respectively. Here $L_{\rm x-tube}$ and $L_{\rm z-tube}$ are the angular momentums from the x-tube and z-tube orbits, respectively. $L_{\rm total}^2=L_{\rm x-tube}^2+L_{\rm z-tube}^2$. 
 In each panel, the x-axis is the radius $r/r_h$ and y-axis is the angular momentum fractions. The x-scale and y-scale run linearly from 0 to 1.}
\label{fig:Lxyz_type}
\end{figure}

Figure~\ref{fig:Lxyz_type} shows the dependence of the angular momentum fractions of x-tube and z-tube orbits on the axial ratio and radius of galaxies. It is noted that z-tube orbits carry most angular momentum at large radius for oblate-triaxial galaxies. For prolate-triaxial galaxies, most angular momentum is from the x-tube orbits. These results are consistent with those of \cite{1994MNRAS.271..924A}.

\begin{figure}
\includegraphics[angle=0, width=80mm]{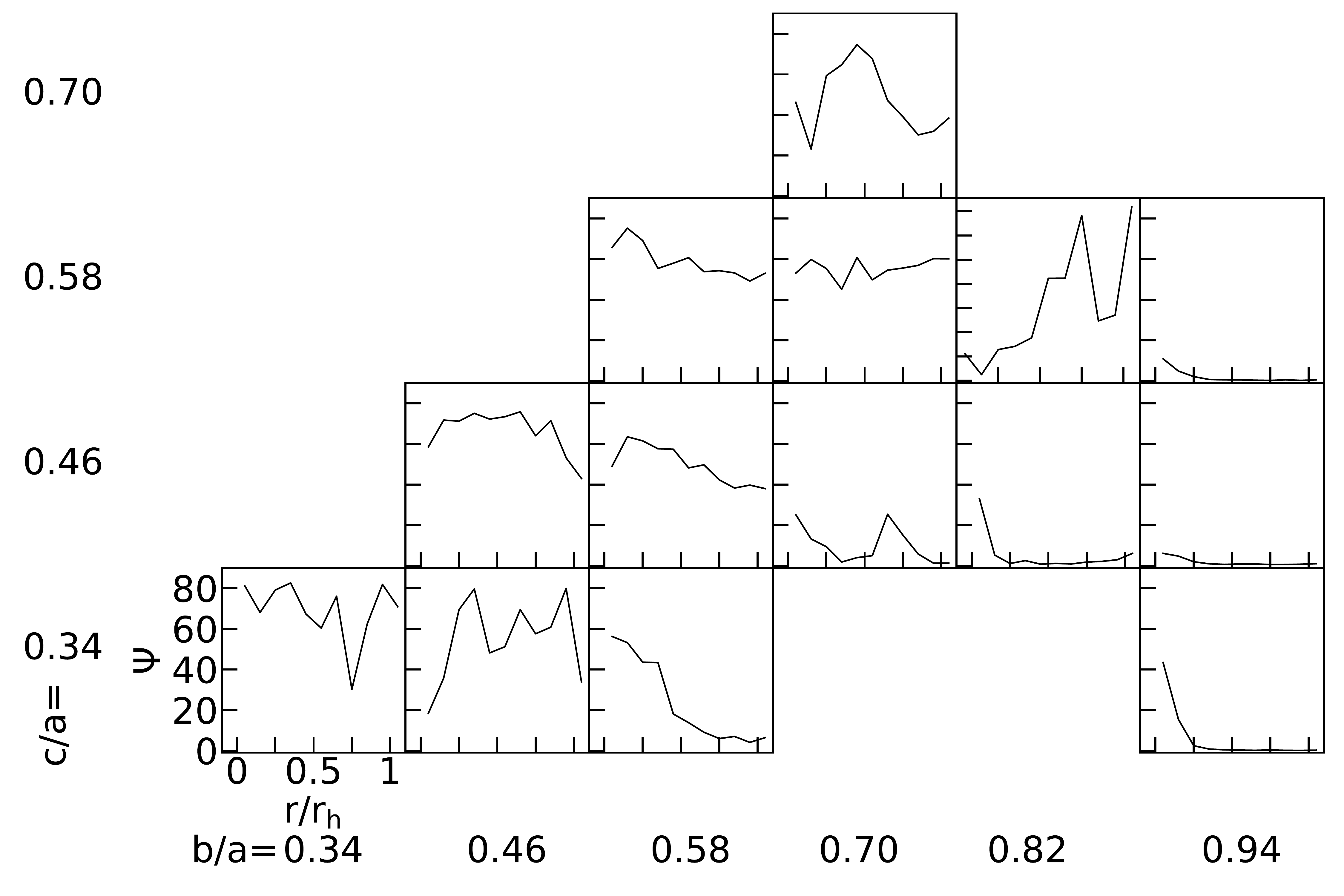}
\caption{Dependence of the misalignment angle $\Psi$ on the axial ratio and radius for all galaxies in our sample. In each panel, the x-scale runs linearly for $r/r_h$ from 0 to 1 and y-scale is from $0$ to $90^{\circ}$.}
\label{fig:Lxyz_ang}
\end{figure}

We follow \cite{1994MNRAS.271..924A} to define the internal misalignment angle $\Psi$ as the angle between the total angular momentum $L_{\rm total}$ and the z-axis by
\begin{equation}
\Psi=\tan^{-1}\bigg(\frac{L_{\rm x-tube}}{L_{\rm z-tube}}\bigg)
\end{equation}
$\Psi=0^{\circ}$ and $\Psi=90^{\circ}$ mean that the angular momentum is along the z-axis and x-axis, respectively. Figure~\ref{fig:Lxyz_ang} displays the dependence of the misalignment angle $\Psi$ on the axial ratio and the radius for all galaxies in our sample.
For the oblate-triaxial systems, the misalignment is close to zero, which indicates the z-tube orbits dominate the angular momentum contributions, especially in the outer region of the galaxies. For the prolate-triaxial galaxies, the angular momentum is dominated by the x-tube orbits. All these results are consistent with those found by \cite{1994MNRAS.271..924A}.

\begin{figure}
\includegraphics[width=\columnwidth]{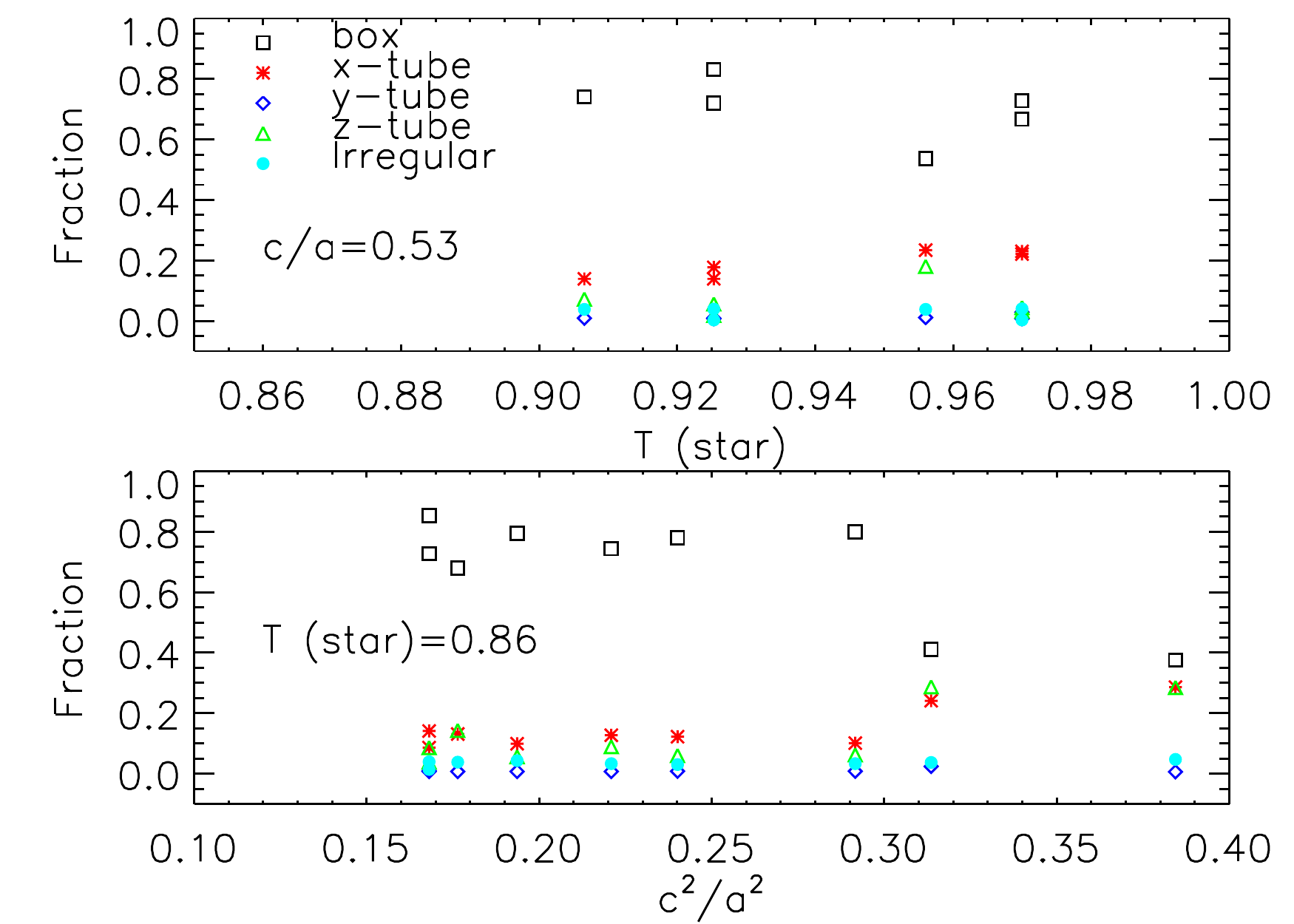}
\caption{Dependence of the orbit fraction within the half stellar mass radius on the triaxiality of the star (up) and the square of the axis ratio $c^2/a^2$ (bottom) for the prolate-triaxial galaxies. The triaxiality is also calculated using the particles within the half stellar mass radius.  The square, star, diamond, triangle and  filled circle symbols represent the orbit fraction for the box, x-tube, y-tube, z-tube and irregular orbits, respectively.}
\label{fig:type_T1}
\end{figure}

Figure~\ref{fig:type_T1} shows the correlation between the fraction of orbit families within the half stellar mass radius and the triaxiality of the stars (up) and  the square of the axis ratio $c^2/a^2$ (bottom) for the prolate-triaxial galaxies. 
It is seen that the fraction of box orbits decreases with the triaxiality of the stars, while the fraction of x-tube orbits increases with the triaxiality of the stars if the axis ratio of the galaxy is fixed.  The population of box orbits decreases with increasing $c^2/a^2$ while the population of  x-tube and z-tube orbits increases with increasing $c^2/a^2$ if $c^2/a^2$ is smaller than 0.4. These results are consistent with those of \cite{1992ApJ...389...79H}.


We follow \cite{2018MNRAS.473.3000Z} and use the circularity $\lambda_i$ to describe different orbit types. The circularity $\lambda_i$ is defined as
\begin{equation}
\lambda_i=\overline{L_i}/(\overline{r}\times\overline{V_c^\prime}) \ \ \  (i=x,y,z)
\end{equation}
where $L_x=yv_z-zv_y$, $L_y=zv_x-xv_z$, $L_z=xv_y-yv_x$, $r=\sqrt{x^2+y^2+z^2}$, and $V_c^\prime=\left|v_x+v_y+v_z\right|$. $\lambda_i=1$ denotes a circular orbit, while $\lambda_i=0$ indicates a box or a radial orbit. 

\begin{figure}
\includegraphics[angle=0, width=90mm]{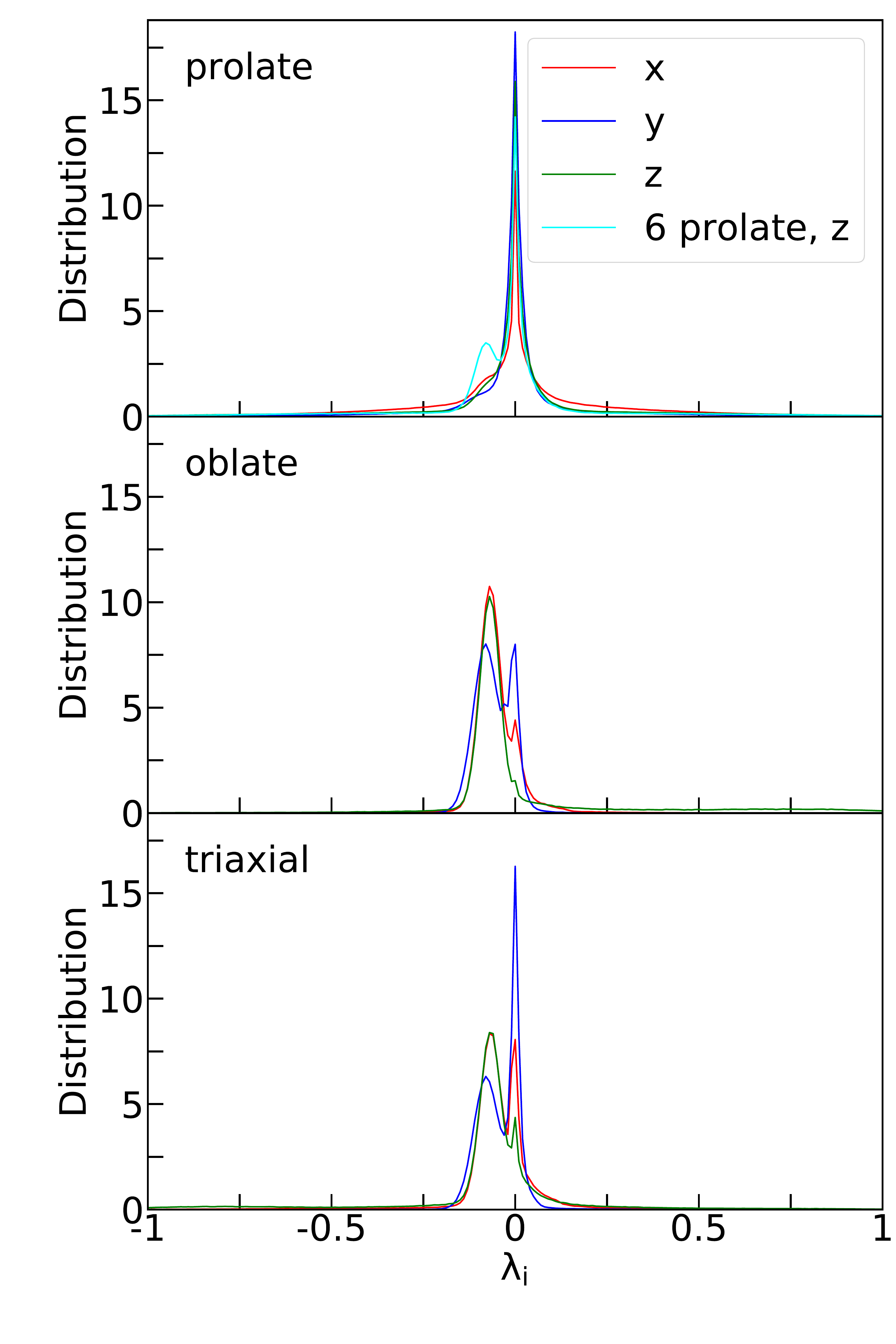}
\caption{Distribution of the circularity $\lambda_x$ (red), $\lambda_y$ (blue), and $\lambda_z$ (green) for the prolate-triaxial (upper), oblate-triaxial (middle) and triaxial galaxies (bottom), respectively. The cyan line in the upper panel represents the result of $\lambda_z$ for 6 randomly selected prolate-triaxial galaxies.}
\label{fig:spin_dis}
\end{figure}

Figure~\ref{fig:spin_dis} shows the average circularity distributions for 35 prolate-triaxial galaxies, 6 oblate-triaxial and 6 triaxial galaxies with the same weight for each galaxy. It is noted that there is a strong peak around $\lambda_i=0$, which indicates that the box or radial orbits dominate in all galaxies. We also find that $\lambda_x$ has a slightly broader distribution than $\lambda_y$ and $\lambda_z$ in the prolate-triaxial galaxies. For the prolate-triaxial galaxies, we find one peak in the circularity distribution. For the oblate-triaxial and triaxial galaxies, there are two peaks, one is at $\lambda_i=0$, and the other is close to $\lambda_i=0$. We have checked all galaxies and found only one peak in the $\lambda_i$ distribution for any single galaxy studied here. The only difference is that the peak is at $\lambda_i=0$ for some galaxies and the peak is close to $\lambda_i=0$ for the remaining ones. If we randomly select 6 prolate-triaxial galaxies, then a peak at $\lambda_i\simeq-0.07$ also appears in distributions of $\lambda_x$, $\lambda_y$ and $\lambda_z$.  Moreover, it is noted that the fraction of the circular orbits is small in all galaxies. 

We also checked the relation between the fraction of the x-, y- and z-tube orbits and the ratio of $\left|L_x\right|/L$, $\left|L_y\right|/L$ and  $\left|L_z\right|/L$ of the dark matter halos within the half stellar mass radius, respectively. 
It is seen that there is no correlation between these two parameters.

\begin{figure}
\includegraphics[angle=0, width=90mm]{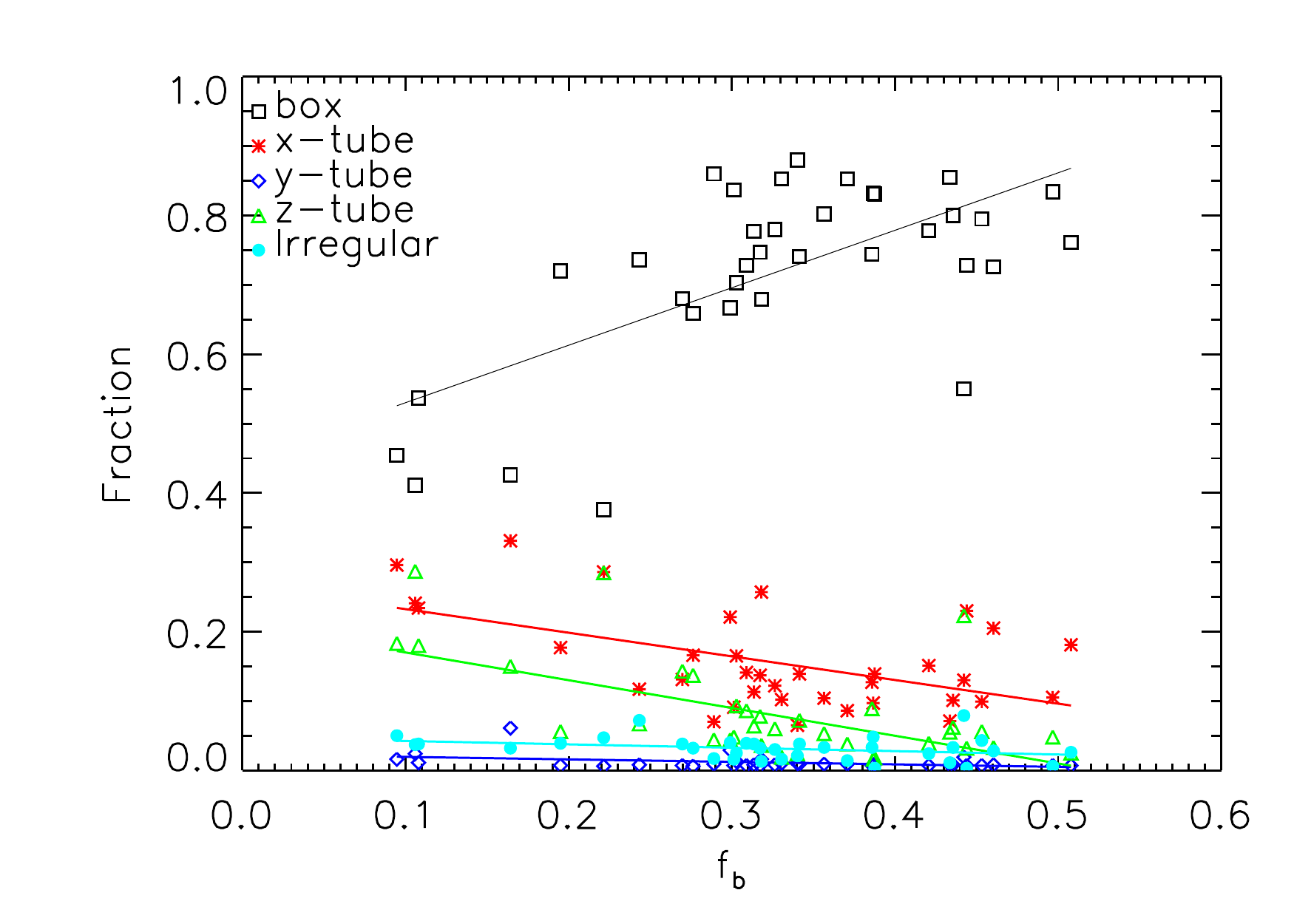}
\caption{Relations between the baryon fraction and the orbit type fraction within the half stellar mass radius. The square, star, diamond, triangle and  filled circle symbols represent the results for the box, x-tube, y-tube, z-tube and irregular orbits, respectively. 
The black, red, blue, green and cyan lines are fits of $Y=A+BX$ for the box, x-tube, y-tube, z-tube and irregular orbits, respectively.}
\label{fig:fb_type}
\end{figure}

Figure~\ref{fig:fb_type} shows the relation between the baryon fraction $f_b$ and the orbit type fraction within the half stellar mass radius. We find a weak correlation between the box orbit faction and the baryon fraction. The fraction of  box orbits increases with increasing baryon fraction, and the Pearson correlation coefficient is 0.65. There is also a weak anti-correlation between the fraction of x-tube orbits and the baryon fraction, which is also found for the z-tube orbits. The Pearson correlation coefficients for x- and z-tube orbits are -0.53 and -0.60, respectively.

\begin{figure}
\includegraphics[angle=0, width=90mm]{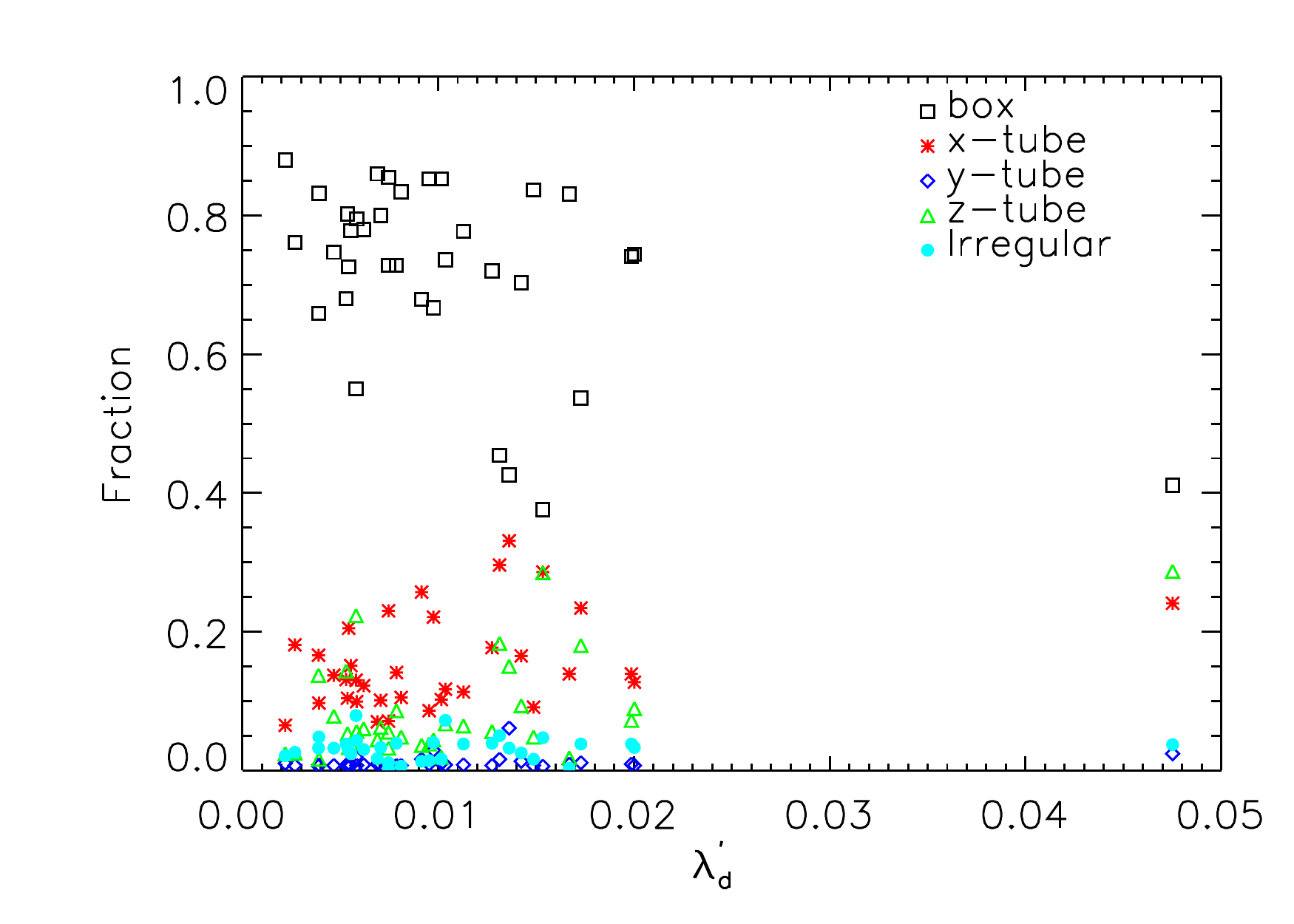}
\caption{Relations between the spin parameter of the dark matter halo and the orbit type fraction within the half stellar mass radius. The square, star, diamond, triangle and  filled circle symbols represent the results for the box, x-tube, y-tube, z-tube and irregular orbits, respectively.}
\label{fig:spin_type}
\end{figure}

Figure~\ref{fig:spin_type} shows the relation between the spin parameter of the dark matter halo $\lambda^{\prime}_d$ and the orbit type fraction within the half stellar mass radius. We find that there is no significant correlation between the fraction of different orbit families and the spin parameter of the dark matter halo.

 \begin{table}
\caption{Spin parameters and the angle between the direction of the angular momentum of the dark matter and that of the stars for 35 prolate-triaxial galaxies, six oblate-triaxial galaxies and six triaxial galaxies (labelled in bold face).  Column (1): The SUBFIND ID at the $z=0$ snapshot. Column (2): Spin parameter for stars within the half stellar mass radius.
Column (3): Spin parameters for dark matter within the half stellar mass radius. Column (4): Spin parameter for considering both the star and dark matter within the half stellar mass radius. Column (5): The angle between the direction of the angular momentum of the dark matter and that of the stars in degrees.}\label{table:spin}
\begin{center}
\begin{tabular}{cccccccllll}\hline
subhalo ID &$\lambda^{\prime}_s$&$\lambda^{\prime}_d$&$\lambda^{\prime}$&$\theta_L$ \\   
(1) & (2) & (3) &(4)&(5)\\
\hline\hline    
 0           &0.0015    &      0.0136   &      0.0150 &30.0 \\
123773   & 0.0049   &     0.0098&       0.0118& 77.7\\
129770   &  0.0016  &     0.0142   &      0.0157 &20.9\\
129771   &  0.0094&       0.0081   &      0.0162 &44.2 \\
132700    &  0.0046   &   0.0075     &    0.0118  &23.6 \\
 135289   &  0.0035   &   0.0079    &      0.0101 &59.7 \\
  138413   &  0.0048 &     0.0058    &      0.0090 &64.7\\
  152864  &   0.0037 &     0.0047  &      0.0074 &55.6 \\
  163932  &    0.0022 &    0.0113     &     0.0134 &14.5 \\
   165890 &    0.0016 &        0.0104    &      0.0112 &64.3 \\
    16937  &    0.0018 &         0.0475    &      0.0491 &29.3 \\
   177128 &        0.0027  &        0.0128    &      0.0153 &22.2 \\
   178998 &        0.0081&        0.0198    &       0.0279 &8.0 \\
    183683 &          0.0016&          0.0062  &        0.0078 &10.5 \\
     185229  &      0.0111 &        0.0095   &       0.0193&41.2 \\
     186924  &      0.0026  &        0.0092    &      0.0115 &28.2 \\
      192506  &     0.0022 &        0.0054    &      0.0075 &19.5\\
       196773 &     0.0047   &        0.0071     &     0.0116&16.7 \\
       200653 &    0.0006  &        0.0058    &     0.0062&50.1 \\
     210738   &  0.0175      &  0.0054    &      0.0229 &4.7\\
     217716   &    0.0022   &   0.0027   &       0.0048 &15.5 \\
     222715   &    0.0023 &     0.0075   &       0.0096 &18.3 \\
     225517   &     0.0106   &   0.0055 &      0.0143 &59.0 \\
     245939    &     0.0033  &        0.0022  &        0.0042 &82.0 &\\
      249937   &     0.0022  &       0.0149   &         0.0169 &23.2 \\
      271246    &   0.0025  &        0.0069   &       0.0081&68.2 \\
       277529   &    0.0010  &        0.0167 &        0.0266 &9.4 \\
       294574   &        0.0020  &        0.0039  &       0.0046 &85.3 \\
       30430    &       0.0029   &       0.0153    &      0.0180&21.2 \\
       324170   &        0.0041  &        0.0102   &       0.0140&24.6\\
       41088    &        0.0006   &        0.0131   &       0.0135 &53.0 \\
       51811    &        0.0022  &         0.0173    &      0.0194 &14.8 \\
       59384    &       0.0010   &       0.0039    &      0.0048 &34.7 \\
        66080   &        0.0020  &        0.0053  &        0.0072 &18.7 \\
         73663   &       0.0091   &       0.0200  &         0.0286 &21.7 \\
         \hline
       \bf{110569}   & 0.0564          &0.0329         &0.0893         &1.5\\     
       \bf{2}             & 0.0606         &0.0101          &0.0702                    &19.2\\       
       \bf{213907}   & 0.0162        &0.0229   &       0.0390 &7.7\\
       \bf{263115}   & 0.0457         &0.0427     & 0.0883                   & 4.0\\      
       \bf{269276}   & 0.0671    &  0.0359 &   0.1029 &1.3 \\
       \bf{73666}     &0.0751    &0.0190    &0.0941                    & 4.5\\  
       \hline
      \bf{127228}    &       0.0122    &       0.0294   &       0.0415  &1.8 \\
       \bf{144528}   &    0.0092       &0.0072          &0.0160    &   27.7 \\
     \bf{154948}     &      0.0147     &     0.0212     &     0.0357    &10.7\\
       \bf{206715}   &  0.0107         &0.0015          &0.0118                    &49.7 \\
        \bf{251546}   &  0.0322                    &  0.0400                   &     0.072               &3.2\\
        \bf{267211}   &  0.0436                    &   0.0145                  &     0.0581               &6.0\\

 \hline  
 
   \hline
 \end{tabular}
\end{center}

\end{table}





\section{summary and discussion}
\label{sec:summary}

The orbit properties are important for understanding the intrinsic structure of prolate-triaxial galaxies. In this paper, we study the orbit properties of 35 prolate-triaxial galaxies first analysed by Li2018  and taken from the Illustris cosmological hydrodynamic simulation. In addition, we also selected six oblate-triaxial galaxies and six triaxial galaxies, which share the same mass and spin ranges as their prolate counterparts. The main results of this paper can be summarized as follows.

1. The spin parameters of the star and dark matter halo in all prolate galaxies are small. Circularities for most orbits in prolate galaxies are close to 0, which indicates box or radial orbits dominate the orbit structure of prolate systems. The spin parameters for stars in the oblate-triaxial galaxies are larger than those in the prolate and triaxial galaxies.

2. The z-tube orbits carry most angular momentum at large radius for the oblate-triaxial galaxies. For prolate galaxies, most angular momentum is from x-tube orbits.

3. Box orbits are found to dominate the orbital structure for most prolate-triaxial galaxies, especially in the central region. Both the x- and z-tubes are important in prolate-triaxial systems, which is different from previous studies that suggested that the z-tube orbits are less important in more prolate systems \citep[e.g.][]{2016ApJ...818..141V}. The fraction of box orbits decreases with increasing galaxy radius, while the fraction of irregular orbits becomes larger with increasing galaxy radius for the prolate-triaxial galaxies. The fraction of x-tube orbits in both oblate-triaxial and triaxial galaxies is smaller than that in the prolate-triaxial galaxies.  
 
4. The fraction of box orbits for prolate-triaxial galaxies ($0.7<T<1$) decreases with the triaxiality, while the fraction of x-tube orbits increases for a given axis ratio $c/a$.

5. There is a weak correlation between the fraction of box orbits and the baryon fraction. The fraction of box orbits increases with increasing baryon fraction, while there is a weak anticorrelation between the fraction of x-tube (or z-tube) orbits and the baryon fraction. 

6. There is no correlation  between the fraction of different orbit families and the spin parameter of the dark matter halo.

7. For massive galaxies, prolate-triaxial, oblate-triaxial or triaxial galaxies, there is only one peak in the distribution of the circularity $\lambda_i$ in a single galaxy. The position of the peak is at $\lambda_i\simeq0$, which indicates that the box (or radial ) orbits dominate the orbit population in these massive galaxies.


It seems that it is impossible for fast rotators to be prolate-triaxial galaxies (also see the upper right panel of Figure 13 in Li2018),  which is also supported by observations \citep{1982Natur.298..728R}.  With the two-dimensional integral field unit spectroscopy of the stellar 
kinematics being carried out, many kinematical data can be used in the future as qualitative constraints in modeling prolate galaxies with the Schwarzschild or Made-to-measure methods. The large scatter between the triaxiality of stars and that of dark matter haloes  increases the complexity in constructing dynamical models with observational data.


\section*{Acknowledgements}
We thank the referee for comments and suggestions that improved the paper. 
Most of the computing was performed on the high performance computing cluster at the Information and Computing Center
at National Astronomical Observatories, Chinese Academy of  Sciences. 
We also acknowledge the support by the National Science
Foundation of China (Grant No. 11821303, 11773034, 11390372, 11633004, QYZDJ-SSW-SLH017), and grants 11333003, 11390372 and 11761131004 to SM. 
SM is also supported by the National Key Basic Research and Development Program of China (No. 2018YFA0404501). VS acknowledges support through the 
Deutsche Forschungsgemeinschaft DFG through project SP 709/5-1.

\bibliographystyle{mn2e}
\bibliography{ms}

\appendix
\section{Galaxies properties and the relative abundance of the orbit familes}
\label{AppendixA}
We present here the galaxy mass, the number of orbits, shapes for both stars and dark matter, and the relative abundance of the orbit families in each galaxy. The detailed information can be found in Table~\ref{table:of_all}.
 \begin{table*}
\caption{Orbit families in 35 prolate-triaxial galaxies, six oblate-triaxial galaxies and six triaxial galaxies (labelled in bold face). Column (1): The SUBFIND ID at the final snapshot. Column (2): Number of orbits in each galaxy.
Column (3): Axis ratios of the stars within the half stellar mass radius. Column (4): Axis ratios of the dark matter within the half stellar mass radius. Column (5): Axis ratios  of the dark matter within the full galaxy. Columns (6)-(10) : The orbit fractions in the full galaxy (within the half stellar mass radius) for the box, x-tube, y-tube, z-tube and irregular orbits, respectively.}\label{table:of_all}
\begin{center}
\begin{tabular}{cclllllllll}\hline
subhalo ID &$\log M_{c200}$& orbit number & axis ratio & axis ratio   &axis ratio &box     &x-tube & y-tube & z-tube &irregular     \\  
     & $(M_{\odot})$  &     & (star $r<r_h$) & (dark $r<r_h$) & (dark all) &$\%$ &$\%$ &$\%$ &$\%$ &$\%$\\ 
(1) & (2) & (3) &(4)&(5)&(6)&(7)&(8)&(9)&(10)&(11)\\
\hline\hline    
 0            & 14.37  &406662&1:0.63:0.60&1:0.70:0.67&1:0.48:0.40& 26.3(42.6)&28.3(33.1)&3.9(6.1)&33.0(15.0)&8.5(3.2)\\
123773   &  13.77  &117589&1:0.55:0.53&1:0.71:0.65&1:0.66:0.55 &46.6(66.7)&22.5(22.1)&16.1(2.9)&9.4(4.4)&5.4(4.0)\\       
 129770   &13.57  & 129307&1:0.69:0.60 &1:0.83:0.79&1:0.51:0.42&50.2(70.3)&20.9(16.5)&3.7(1.3)&21.0(9.3)&4.2(2.5) \\   
 129771   &13.57  & 59860 &1:0.68:0.56&1:0.76:0.71 &1:0.82:0.76& 62.3(83.4) &18.7(10.5) &3.1(0.7)& 14.3(4.8) & 1.5(0.6)\\   
 132700  &11.76  &72089&1:0.62:0.46&1:0.71:0.63 &1:0.92:0.72&70.3(85.5)&11.5(7.1)&0.6(0.8)&16.0(5.5)&1.6(1.1) \\   
 135289   &13.56  &73673&1:0.52:0.41&1:0.80:0.72&1:0.74:0.59 &48.2(72.8)&16.3(14.1)&0.7(0.7)&28.2(8.6)&6.5(3.9)\\
 138413   &13.77  &134709&1:0.50:0.46&1:0.82:0.78 &1:0.58:0.47&55.0(55.0)&13.0(13.0)&1.9(19)&22.3(22.3)&7.9(7.9)\\       
  152864   &13.63 &108456&1:0.58:0.47&1:0.65:0.58 &1:0.82:0.65 & 62.7(74.7)&15.3(13.7)&0.7(0.7)&18.6(7.8)&2.7(3.2)\\
  163932   &13.46   &89378&1:0.55:0.46&1:0.66:0.59&1:0.69:0.64& 62.2(77.7)&19.6(11.3)&0.7(0.8)&13.9(6.4)&3.7(3.8)\\
  165890   &13.62  & 77671&1:0.53:0.45&1:0.84:0.79&1:0.64:0.45&58.0(73.6)&10.9(11.7)&0.7(0.8)&22.7(6.7)&7.6(7.2)\\
  16937     &14.35  &276946&1:0.63:0.56&1:0.74:0.66&1:0.56:0.41&24.8(41.1)&17.9(24.1)&1.8(2.4)&42.8(28.7)&12.7(3.7) \\
  177128   &13.41  &44334&1:0.57:0.52&1:0.81:0.77&1:0.59:0.49 & 51.5(72.0)&20.1(17.7)&0.8(0.7)&19.0(5.6)&8.5(3.9)\\    
  178998    &13.62 &87351&1:0.59:0.53 &1:0.77:0.72&1: 0.63:0.50&53.6(74.1)&13.4(13.9)&4.3(0.9)&22.4(7.2)&6.3(3.8)\\  
  183683    &13.51  &59169&1:0.58:0.49&1:0.71:0.62&1:0.90:0.74& 62.6(78.0)&10.5(12.2)&0.7(0.8)&20.2(6.0)&6.0(3.0)\\ 
  185229   &13.08  &57749&1:0.52:0.41&1:0.57:0.51&1:0.89:0.77 &65.5(85.3)&17.2(8.6)&0.7(0.9)&15.3(3.8)&1.4(1.4)\\
  186924   &13.51    & 142724&1:0.66:0.64&1:0.80:0.80&1:0.81:0.76&50.0(67.9)&36.0(25.7)&8.6(1.6)&4.1(3.6)&1.3(1.3)\\
  192506    &13.43   &74193&1:0.49:0.41 &1:0.72:0.62&1:0.63:0.43&70.2(80.2)&9.7(10.4)&0.7(0.9)&16.2(5.3)&3.0(3.3)\\
  196773    &13.38     &74901&1:0.62:0.54&1:0.76:0.72&1:0.84:0.64 &69.0(80.0)&10.2(10.1)&0.8(0.8)&15.6(6.2)&4.3(3.3)\\
  200653    &13.34     &56486&1:0.54:0.44&1:0.71:0.60&1:0.58:0.54&68.1(79.5)&9.1(9.9)&0.6(0.7)&16.7(5.6)&5.5(4.3)\\
  210738    &13.14    &67507&1:0.44:0.43&1:0.70:0.67 &1:0.78:0.75 & 62.8(72.6)&22.2(20.5)&2.5(0.8)&10.0(3.3)&2.5(2.8)\\
  217716    &13.22    &60062&1:0.45:0.41&1:0.71:0.69&1:0.79:0.63&  60.1(76.1)&24.4(18.1)&0.7(0.7)&12.1(2.5)&2.8(2.6)\\
  222715    &13.07    &65549&1:0.55:0.53&1:0.70:0.69&1:0.82:0.69  & 59.2(72.8)&24.9(23.0)&0.5(0.7)&13.3(3.2)&2.1(0.3)\\
  225517    &13.20  &38947&1:0.54:0.54&1:0.83:0.77&1:0.88:0.63&68.1(77.8)&12.0(15.1)&0.7(0.7)&16.1(3.9)&3.1(2.4)\\  
  245939    &13.06   &45260&1:0.49:0.38&1:0.57:0.51&1:0.81:0.71 & 74.6(88.0)&13.1(6.5)&0.8(1.0)&9.6(2.4)&2.0(2.1)\\  
  249937     &13.05   &34098&1:0.70:0.62 &1:0.73:0.68 &1:0.55:0.48 &71.8(83.7)&14.0(9.1)&0.7(0.7)&11.0(4.8)&2.5(1.6)\\      
  271246     &13.03   &35951&1:0.55:0.47&1:0.65:0.56&1:0.79:0.70&69.2(86.0)&8.6(7.0)&0.7(0.9)&19.9(4.4)&1.5(1.7)\\
  277529    &12.78   &33231&1:0.57:0.52&1:0.60:0.59 &1:0.74:0.69  &65.3(83.1)&26.3(13.9)&0.7(0.9)&6.5(1.8)&1.2(0.3)\\
  294574    &12.82    &21897&1:0.35:0.31 &1:0.68:0.63&1:0.70:0.65&70.6(83.2)&17.0(9.7)&2.8(0.8)&6.2(1.6)&3.5(4.8)\\
  30430      &14.34    &377515&1:0.68:0.62&1:0.80:0.71 &1:0.66:0.57 &34.1(37.6)&23.3(28.6)&0.8(0.6)&35.6(28.5)&6.2(4.7)\\
  324170   &12.66  &23187&1:0.65:0.61 &1:0.64:0.61 &1:0.68:0.61&  76.3(85.3) &13.5(10.2) &0.7(1.0) & 7.3(1.9)&2.1(1.6) \\   
  41088     &14.07  &177309&1:0.73:0.71&1:0.77:0.74 &1:0.72:0.59 &28.1(45.4)&29.1(29.6)&1.2(1.6)&33.6(18.3)&8.2(5.0)\\
  51811     &14.23  &146288&1:0.55:0.52 &1:0.77:0.72   &1:0.70:0.63  & 34.6(53.7)&19.2(23.4)&1.1(1.1)&29.9(18.0)&15.2(3.8)\\
  59384    &14.11   &326687&1:0.59:0.49&1:0.74:0.64&1:0.65:0.50&44.9(65.9)&18.8(16.6)&0.6(0.6)&30.9(13.7)&4.7(3.2)\\
  66080    &14.13   &331196&1:0.53:0.42&1:0.66:0.56&1:0.62:0.45 &46.3(68.0)&16.3(13.1)&0.7(0.7)&30.9(14.3)&5.7(3.8) \\ 
 73663     &13.64  &161692&1:0.57:0.47&1:0.68:0.61&1:0.30:0.27&51.4(74.4)&17.7(12.7)&0.7(0.7)&23.3(8.9)&6.8(3.3)\\
 \hline
 \bf{110569}  & 13.49 &  54561  &  1:1:0.62 &1:0.98:0.81&1:0.99:0.71&20.8(33.7)&0.6(0.8)&0.5(0.4)&77.4(64.9)&0.7(0.2)\\ 
 \bf{2}            &14.37  & 51473   & 1:0.95:0.33 &1:0.95:0.82&1:0.91:0.64 &49.4(61.5)& 1.5(0.7)& 2.2(0.7) &43.7(35.5) &3.1(1.6)   \\
  \bf{213907}  &13.25 & 49307    &1:0.88:0.51     &1:0.93:0.72         &1:0.61:0.47 &59.1(72.8)&6.4(3.5)&0.9(0.7)&28.1(21.2)&5.5(1.7)\\
 \bf{263115}   &12.96 &41016 & 1:0.96:0.51&1:0.96:0.71 &1:0.95:0.51&55.2(74.4)&2.2(1.2)&0.6(0.8)&40.4(22.9)&1.7(0.7)     \\
 \bf{269276}  &12.93 & 58803 &  1:0.91:0.56                &1:0.97:0.81                 &1:0.88:0.80 &40.4(62.2) &1.5(2.3)&1.3(0.7)&55.5(34.4)&1.3(0.3) \\
 \bf{73666}    &13.64 &70614&1:0.96:0.41 &1:0.91:0.70&1:0.92:0.51 & 43.2(59.1) &4.5(1.6) &0.4(0.6)& 49.8(38.3)& 2.0(0.4) \\ 
 \hline
 \bf{127228}   &13.80 &158988            &1:0.69:0.43                  &1:0.77:0.54               &1:0.91:0.66&54.4(76.9) &7.2(6.0)&0.7(0.8)&33.7(12.9)&3.9(3.4) \\
 \bf{144528}   &13.51 &112345&1:0.78:0.58&1:0.79:0.73&1:0.62:0.53&57.6(74.0) &17.5(8.4) &0.7(0.7)&19.4(12.5)&4.7(4.3)  \\
 \bf{154948}   &13.40 &97955             &1:0.70:0.58                  &1:0.82:0.66               &1:0.66:0.59&65.9(80.6)&8.6(7.6)&0.6(0.8)&21.9(8.9)&2.9(2.1)\\      
 \bf{206715}   &12.87 &48938&1:0.60:0.37&1:0.69:0.61  &1:0.72:0.47 &61.1(80.5) &11.5(7.2) &0.6(0.9)&26.1(11.3) &0.6(0.1)  \\
 \bf{251546}   &13.10 &67101&1:0.86:0.52 &1:0.93:0.72 &1:0.91:0.51 & 52.0 (76.2)  & 2.3 (2.5) & 0.5(0.8)& 44.4 (20.2)&0.8(0.2)  \\
 \bf{267211}   &13.06 &50475 &1:0.63:0.39&1:0.78:0.62 &1:0.76:0.54 &67.7(88.9) & 10.2(2.7) & 0.8(1.1) & 19.7(6.6) &1.5 (0.7)  \\

 \hline  
 
   \hline
 \end{tabular}
\end{center}

\end{table*}

\section{Orbit classification by using the angular momentum}
\label{AppendixB}
The AM method is based on the values of the three angular momentum of orbits. We follow~\cite{2008MNRAS.385..647V} and~\cite{2017ApJ...844..130W} to classify the orbits by the sign of the maximum angular momentum $\ max(L_i)$ and the minimum angular momentum
$min(L_i)\! (i=x,y,z)$ . The box orbits are defined as

\begin{eqnarray}
\rm max(L_x)\times min(L_x)<0\\
\rm max(L_y)\times min(L_z)<0\\
\rm max(L_z)\times min(L_z)<0
\end{eqnarray}

The x-tube orbits are 
\begin{eqnarray}
\rm max(L_x)\times min(L_x)>0\\
\rm max(L_y)\times min(L_z)<0\\
\rm max(L_z)\times min(L_z)<0
\end{eqnarray}

The y-tube orbits are 
\begin{eqnarray}
\rm max(L_x)\times min(L_x)<0\\
\rm max(L_y)\times min(L_z)>0\\
\rm max(L_z)\times min(L_z)<0
\end{eqnarray}

The z-tube orbits are
z-tube
\begin{eqnarray}
\rm max(L_x)\times min(L_x)<0\\
\rm max(L_y)\times min(L_z)<0\\
\rm max(L_z)\times min(L_z)>0
\end{eqnarray}

The remainder orbits are classified as the irregular orbits.

 \begin{table*}
\caption{Orbit families in 35 prolate-triaxial galaxies, six oblate-triaxial galaxies and six triaxial galaxies (labelled in bold face). Column (1): The SUBFIND ID at the final snapshot. Column (2): Triaxiality of the stars within the half stellar mass radius. Column (3): Triaxiality of the dark matter within the half stellar mass radius. Column (4): Triaxiality of the dark matter within the full galaxy. Columns (5)-(9) : The orbit fractions in the full galaxy by using CA/AM method for the box, x-tube, y-tube, z-tube and irregular orbits, respectively. Column (10): Consistent fraction between CA and AM methods.}\label{table:comp}
\begin{center}
\begin{tabular}{cclllllllll}\hline
subhalo ID &T & T& T& box     &x-tube & y-tube & z-tube &irregular & consistent fraction     \\  
     & (star $r<r_h$) & (dark $r<r_h$) & (dark all) & $\%$ &$\%$ &$\%$ &$\%$ &$\%$ &$\%$  \\ 
(1) & (2) & (3) &(4)&(5)&(6)&(7)&(8)&(9)&(10)\\
\hline\hline    
 0            & 0.94   & 0.93   & 0.92 & 26.3/34.4&28.3/34.5&3.9/2.3&33.0/6.0&8.5/22.9 &58.8\\
123773   & 0.97   & 0.86   & 0.81&46.6/50.5&22.5/23.1&16.1/17.5&9.4/0.9&5.4/8.0&78.3\\       
 129770  & 0.82   & 0.83   & 0.90&50.2/61.8&20.9/19.3&3.7/3.0&21.0/0.4&4.2/12.3&68.2\\   
 129771  & 0.78   & 0.85   & 0.78& 62.3/75.4 &18.7/14.0 &3.1/1.7& 14.3/1.4 & 1.5/7.6 &73.1\\   
 132700  & 0.78   & 0.82   & 0.32&70.3/76.0&11.5/9.4&0.6/0.0&16.0/13.1&1.6/1.5 & 89.9 \\   
 135289   & 0.88   & 0.75   & 0.69&48.2/54.9&16.3/16.0&0.7/0.1&28.2/16.8 & 6.5/12.2&74.8\\
 138413   & 0.95   & 0.84   & 0.85&55.0(55.0)&13.0(13.0)&1.9(19)&22.3(22.3)&7.9(7.9)&72.8\\       
  152864   & 0.85   & 0.87   & 0.57&62.7/67.7&15.3/14.1&0.7/0.0&18.6/15.9&2.7/2.3&87.9\\
  163932   & 0.88   & 0.87   & 0.89& 62.2/67.7&19.6/18.9&0.7/0.1&13.9/10.5&3.7/2.9&86.8\\
  165890   & 0.90   & 0.78   & 0.74&58.0/70.6&10.9/8.7&0.7/0.2&22.7/9.2&7.6/11.3&72.3\\
  16937     & 0.88   & 0.80   & 0.83&24.8/27.2&17.9/20.6&1.8/1.6&42.8/17.5&12.7/33.1&62.2\\
  177128   & 0.93   & 0.84   & 0.86&51.5/55.1&20.1/21.4&0.8/0.0&19.0/11.4&8.5/12.1&77.2\\    
  178998   & 0.91   & 0.85   & 0.80&53.6/69.5&13.4/12.2&4.3/2.8&22.4/2.5&6.3/12.9&63.6\\  
  183683    & 0.87   & 0.81   & 0.42&62.6/68.5&10.5/8.4&0.7/0.0&20.2/16.8&6.0/6.2&84.6\\ 
  185229   & 0.88   & 0.91   & 0.51&65.5/72.4&17.2/14.8&0.7/0.0&15.3/10.4&1.4/2.4&86.9\\
  186924    & 0.96   & 1.00   & 0.81&50.0/47.1&36.0/41.9&8.6/8.4&4.1/1.0&1.3/1.5&80.5\\
  192506   & 0.91   & 0.78   & 0.74&70.2/76.1&9.7/7.6&0.7/0.0&16.2/15.2&3.0/1.1&90.1\\
  196773     & 0.87   & 0.88   & 0.50&69.0/75.8&10.2/8.5&0.8/0.1&15.6/10.9&4.3/4.7&85.3\\
  200653    & 0.88   & 0.77   & 0.94&68.1/(79.5)&9.1(9.9)&0.6(0.7)&16.7(5.6)&5.5(4.3)&85.4\\
  210738    & 0.99   & 0.93   & 0.90& 62.8/69.3&22.2/22.8&2.5/2.0&10.0/0.7&2.5/5.2 &73.0\\
  217716    & 0.96   & 0.95   & 0.62& 60.1/64.5&24.4/26.3&0.7/0.0&12.1/3.7&2.8/5.4&75.2\\
  222715   & 0.97   & 0.97   & 0.63& 59.2/64.8&24.9/22.4&0.5/0.0&13.3/10.4&2.1/2.4 &85.0\\
  225517   & 1.00   & 0.76   & 0.37&68.1/77.4&12.0/11.1&0.7/0.0&16.1/6.7&3.1/4.8&77.6\\  
  245939    & 0.89   & 0.91   & 0.69&74.6/81.5&13.1/9.8&0.8/0.0&9.6/7.9&2.0/0.8&90.0\\  
  249937    & 0.83   & 0.87   & 0.91&71.8/76.7&14.0/12.7&0.7/0.0&11.0/8.8&2.5/1.7 & 87.9\\      
  271246    & 0.90   & 0.84   & 0.74 &69.2/75.1&8.6/6.1&0.7/0.0&19.9/17.8&1.5/0.9&91.2\\
  277529     & 0.93   & 0.98   & 0.86&65.3/70.7&26.3/24.4&0.7/0.0&6.5/4.0&1.2/0.9 &84.5\\
  294574    & 0.97   & 0.89   & 0.88 &70.6/83.6&17.0/11.5&2.8/2.0&6.2/0.3&3.5/2.7&78.6\\
  30430      & 0.87   & 0.73   & 0.84 &34.1/35.6&23.3/26.5&0.8/0.2&35.6/20.3&6.2/17.4&73.0\\
  324170   & 0.92   & 0.94   & 0.86 & 76.3/83.7 &13.5/9.6 &0.7/0.0 & 7.3/6.0&2.1/0.6 &85.3 \\   
  41088   & 0.94   & 0.90   & 0.74  &28.1/32.7&29.1/35.0&1.2/0.2&33.6/10.7&8.2/21.4&62.4\\
  51811     & 0.96   & 0.85   & 0.85 & 34.6/40.9&19.2/23.4&1.1/0.4&29.9/8.6&15.2/26.6&52.2\\
  59384     & 0.86   & 0.77   & 0.77&44.9/47.5&18.8/19.7&0.6/0.0&30.9/24.7&4.7/8.0&82.3\\
  66080    & 0.87   & 0.82   & 0.77&46.3/49.4&16.3/16.7&0.7/0.0&30.9/26.3&5.7/7.6 &84.2 \\ 
 73663     & 0.87   & 0.86   & 0.98 &51.4/56.3&17.7/17.4&0.7/0.0&23.3/16.7&6.8/9.6&82.1\\
 \hline
 \bf{110569} & 0.00   & 0.12   & 0.04 &20.8/24.2&0.6/0.5&0.5/0.8&77.4/72.6&0.7/1.9&86.4\\ 
 \bf{2}           & 0.11   & 0.30   & 0.29 &49.4/68.0& 1.5/3.9& 2.2/0.4 &43.7/18.2 &3.1/9.6&64.8 \\
  \bf{213907}  & 0.30   & 0.28   & 0.81 &59.1/63.8&6.4/6.1&0.9/0.0&28.1/22.3&5.5/7.7&83.9\\
 \bf{263115}  & 0.11   & 0.16   & 0.13 &55.2/61.8&2.2/1.5&0.6/0.0&40.4/33.5&1.7/3.1&81.5     \\
 \bf{269276} & 0.25   & 0.17   & 0.63  &40.4/43.5 &1.5/0.9&1.3/1.1&55.5/53.7&1.3/7.3&87.5 \\
 \bf{73666}  & 0.09   & 0.34   & 0.21  & 43.2/50.5 &4.5/3.9 &0.4/0.0& 49.8/43.7& 2.0/1.9&85.5\\ 
 \hline
 \bf{127228}   & 0.64   & 0.57   & 0.30 &54.4/59.0 &7.2/6.4&0.7/0.1&33.7/30.2&3.9/4.4 &86.7\\
 \bf{144528}  & 0.59   & 0.80   & 0.86 &57.6/66.6 &17.5/17.2 &0.7/0.3&19.4/8.5&4.7/7.4 &78.1 \\
 \bf{154948}  & 0.77   & 0.58   & 0.87 &65.9/71.9&8.6/7.5&0.6/0.1&21.9/18.3&2.9/2.3&88.2\\      
 \bf{206715}  & 0.74   & 0.83   & 0.62  &61.1/67.2 &11.5/8.8 &0.6/0.0&26.1/23.7 &0.6/0.3&91.8  \\
 \bf{251546} & 0.36   & 0.28   & 0.23  & 52.0/ 59.2  & 2.3 /1.2 & 0.5/0.0& 44.4/ 38.6&0.8/0.9&87.6  \\
 \bf{267211}  & 0.71   & 0.64   & 0.60  &67.7/73.3 & 10.2/10.2 & 0.8/0.0 & 19.7/14.7 &1.5/1.7 &87.3  \\

 \hline  
 
   \hline
 \end{tabular}
\end{center}

\end{table*}

\label{lastpage}

\clearpage
\end{document}